\documentclass[aps,prr,twocolumn,floatfix,amssymb]{revtex4-1} 
\usepackage[utf8]{inputenc}
\usepackage[english]{babel}
\usepackage{amsmath}
\usepackage{amsfonts}
\usepackage{amssymb}
\usepackage{graphicx}
\usepackage{float}
\usepackage{braket}
\usepackage{ulem}
\usepackage{appendix}
\usepackage[dvipsnames]{xcolor}
\usepackage{xcolor}
\usepackage[breaklinks=true,colorlinks,citecolor=blue,linkcolor=blue,urlcolor=blue]{hyperref}
\newcommand{\be}{\begin{equation}}
\newcommand{\ee}{\end{equation}}
\newcommand{\ber}{\begin{eqnarray}}
\newcommand{\eer}{\end{eqnarray}}
\newcommand{\nn}{\nonumber}

\newcommand{\pv}{{\bf p}}

\newcommand{\jv}{{\bf j}}
\newcommand{\rv}{{\bf r}}
\newcommand{\qv}{{\bf q}}

\def\eer{\end{eqnarray}}

\def\rv{{\bf r}}

\def\pv{{\bf p}}

\def\uv{{\bf u}}

\def\jv{{\bf j}}

\def\qv{{\bf q}}

\def\uv{{\bf u}}

\def\nn{\nonumber}

\begin{document}

\title{
Collective excitations and quantum incompressibility  in electron-hole bilayers
}
\author {S. De Palo$^{1,2}$, P. E. Trevisanutto$^{3,4}$, G. Senatore$^{2}$, and G. Vignale$^{5}$}
\affiliation{$^{1}$ CNR-IOM-DEMOCRITOS, Trieste, Italy}
\affiliation{$^{2}$ Dipartimento di Fisica,
Universit\`a di Trieste, strada Costiera 11, 34151
Trieste, Italy}
\affiliation{$^{3}$ Centre for Advanced 2D Materials, National University of Singapore, 6 Science Drive 2, 117546 Singapore}
\affiliation{$^{4}$ European Centre for Theoretical Studies in Nuclear Physics and Related Areas (ECT*-FBK) and Trento Institute for Fundamental Physics and Applications
(TIFPA-INFN), Via Sommarive, 14, 38123 Povo TN, Trento, Italy}
\affiliation{$^5$Department of Physics and Astronomy, University of Missouri, Columbia, Missouri 65211, USA\\}

\begin{abstract}

We apply quantum continuum mechanics to the calculation of the  excitation spectrum of a coupled electron-hole bilayer. The theory expresses excitation energies in terms of ground state intra- and inter-layer pair correlation functions, which are available from  Quantum Monte Carlo calculations.   The final formulas for the collective modes deduced from this approach coincide with the formulas obtained in the ``quasi-localized particle approximation" by Kalman et al., and likewise the theory predicts the existence of gapped excitations in the charged channels,  with the gap arising from electron-hole correlation.   An immediate consequence of the gap is that the static density-density response function of the charged channel vanishes as $q^2$ for wave vector $q \to 0$, rather than linearly in $q$, as commonly expected. In this sense, the system is {\it incompressible}. This feature, which has no analogue in the classical electron-hole plasma, is consistent with the existence of an excitonic ground state, and implies the existence of a discontinuity in the chemical potential of electrons and holes when the numbers of electrons and holes are equal. It should be experimentally observable by monitoring the densities of electrons and holes in response to potentials that attempt to change these densities in opposite directions.

\end{abstract}

\maketitle

\section{Introduction}

\begin{figure}
\centering
\includegraphics[width=0.4\textwidth]{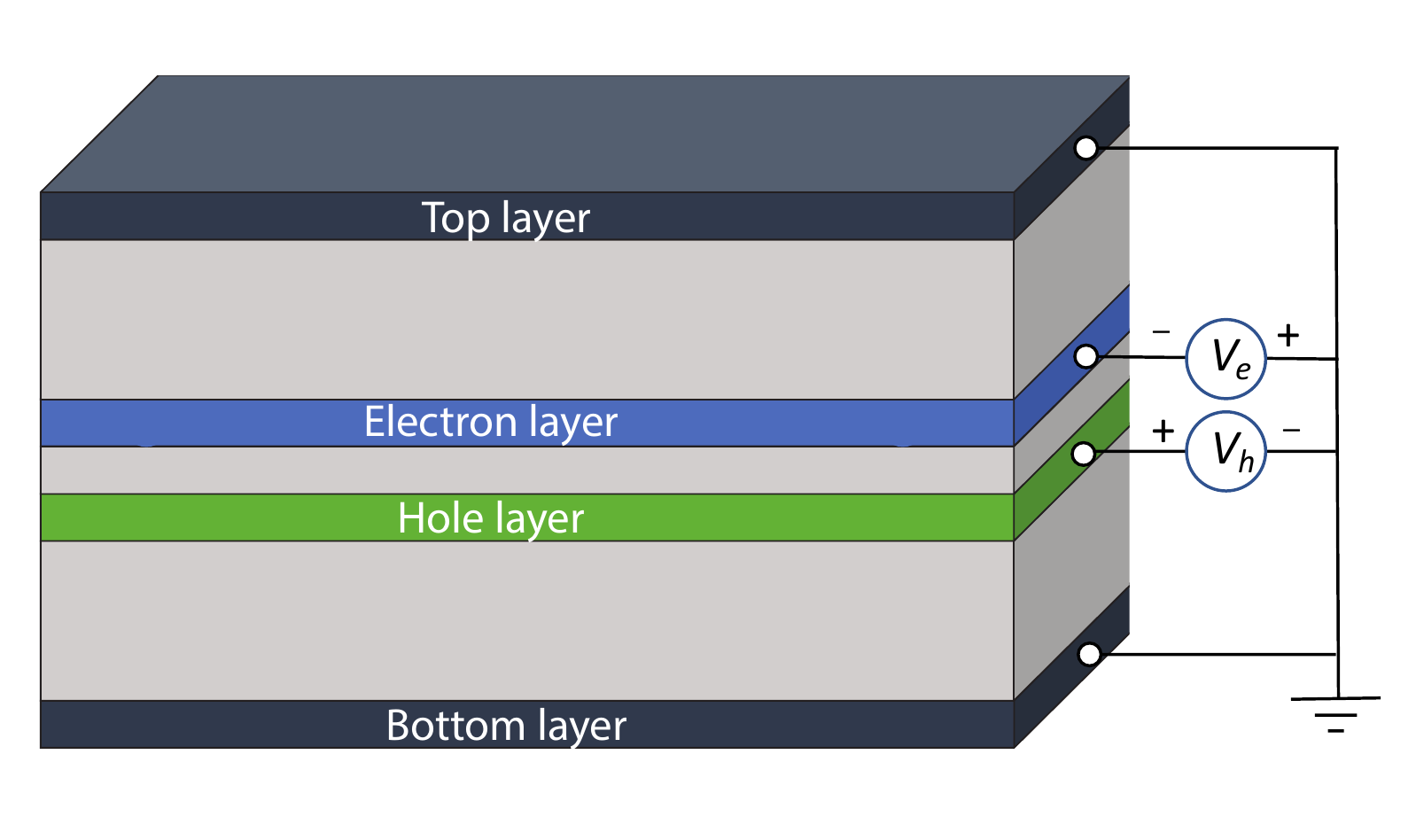}
\caption{Schematics of an electron-hole bilayer heterostructure.  The densities of electrons and holes are controlled by changing the potentials of the electron and hole layers relative to the top and bottom layers (gates).}
\label{fig:Heterostruct}
\end{figure}

Two-dimensional electron-hole systems, in which electrons and holes reside in well separated layers of a semiconductor heterostructure (see Fig.\ref{fig:Heterostruct}), have received much attention in recent years.  
``Well separated" means that tunneling between the layers is negligible.  Voltages $V_e$ and $V_h$ applied to the electron (e) and hole (h) layers respectively can be used to control the carrier densities in each layer.  The attractive interaction between electrons and holes creates a rich phase diagram 
in which BCS pairing at high density, exciton and multi-exciton formation at low density, compete with the  conventional Fermi liquid phase.  

Quantum Monte Carlo (QMC) calculations of the ground state wave function of this system have provided compelling evidence for the existence of non-Fermi liquid states with the formation of exciton and multi-exciton complexes and the appearance of off-diagonal long-range order at low densities \cite{depalo2002,depalo2003,shumway2012,maezono2013,sharma2016,tramonto}.  Experimental signatures of electron-hole pairing and superfluidity  have been seen in counterflow experiments at high magnetic field~\cite{Eisenstein2004,Tutuc2004}, in multi-layer structures of transition metal dichalcogenides~\cite{Conti2020-I,Conti2020-II,Mak2019}, 
and are also expected to emerge from Coulomb drag experiments~\cite{Vignale96}.  In this paper we will use the available information about the ground state properties of the electron-hole bilayer to achieve something that is presently beyond the reach of QMC, namely to predict the density fluctuation spectrum and the transverse current fluctuation spectrum.

We will focus on {\it symmetric} electron-hole systems, meaning that the densities and the effective masses of electrons and holes are identical and the elementary excitations can be classified as ``symmetric" (electrons and holes moving in phase) and ``antisymmetric" (electrons and holes moving out of phase).

The excitation spectrum is largely determined by the character of the ground state.  For example, in the Fermi liquid phase, there are two types of excitations: single-particle excitations out of the Fermi sea, and  collective modes (plasmons), in which the electrons and the holes oscillate out of phase, with a dispersion $\propto \sqrt{q}$, where $q$ is the wave vector.  In principle, an acoustic plasmon mode could also exist, with electrons and hole oscillating in phase,  at sufficiently large interlayer separation~\cite{Santoro1988}.  All these excitations are gapless. The transverse current excitation spectrum is also gapless.

The situation is quite different in electron-hole paired phases.  A fundamental property of the paired ground state is its {\it rigidity} with respect to perturbations that tend to shift electron and hole densities in opposite directions.  This can be understood as follows. Adding (or removing) equal numbers of electrons and holes in a small volume of the system can be viewed as increasing (or decreasing) the number of bound electron-hole pairs without breaking any bond. The sum of the addition and removal energies of a pair is $2\mu-2\mu=0$, where $\mu$ is the chemical potential and $2\mu$ is the energy of a bound electron-hole pair: this is what we mean when we say that this mode of excitation is gapless. In contrast, adding an electron while removing one hole changes the energy by $-2\mu+2\epsilon_e$, where $\epsilon_e$ is the energy of a free (unbound) electron.  Similarly, removing an electron while adding a hole changes the energy by $-2\mu+2\epsilon_h$, where $\epsilon_h$ is the energy of a free (unbound) hole. The sum of the addition and removal energies is now $2(\epsilon_e+\epsilon_h-2\mu)$, which is twice the binding energy of an electron-hole pair: hence, this mode of excitation is gapped.

 In order to calculate the excitation spectra of electron-hole bilayers we resort to  ``quantum continuum mechanics" (QCM -- not to be confused with QMC)  for quantum electronic systems. This theory was discussed in detail in Ref.~\cite{Gao10}.  It yields an exact equation of motion (Eq. (15) in Ref.~\cite{Gao10}) for the current density in terms of a stress tensor field, which is, in principle, a functional of the current density itself. In the ``elastic approximation" (see Ref.~\cite{Gao10}) the equation of motion for the current is obtained by making an Ansatz on the form of the solution of the time-dependent Schr\"odinger equation, namely, that the time-dependent wave function is obtained  by applying a time-dependent deformation to the ground state wave function.  Equivalently, one can say that the wave function remains constant in an accelerated reference frame in which the density is constant and each infinitesimal volume element of the system remains at rest.  Under these assumptions, the equation of motion for the current density can be expressed in terms of ground state properties, such as the momentum occupation number and the pair distribution functions, which can be accurately computed by QMC. This approach is justifiable only for strongly interacting systems, whose dynamics is dominated by collective motions of relatively large groups of electrons (the volume elements of the fluid), while single particle excitations are negligible or absent.   For example $^4$He and electrons in the lowest Landau level~\cite{Girvin1986} are well described by this method.  Low-density electron liquids with strong Wigner crystal-like correlations between the positions of the electrons are expected to be well described too.  Generally, the method works well for collective modes.  The single-particle portion of the excitation spectrum,   if  present, is  absorbed in the collective mode spectrum, in such a way that certain spectral sum rules (f-sum rule, third moment sum rule) are satisfied. 
 
 In this paper we apply QCM to the calculation of the excitation spectrum of an electron hole bilayer.  Remarkably, the final expressions for the collective mode frequencies coincide with the expressions that were obtained by Kalman et al. \cite{kalman-qlca} by using what they called  the ``quasi-localized particle approximation" (QLCA).   Thus, we can say that  QCM offers a way to formalize the physical assumptions underlying the QLCA. 
 
 The most interesting result of the calculation is that the antisymmetric sector of the spectrum is gapped.  Thus, in contrast to the Fermi liquid, where the antisymmetric mode (electrons and holes oscillating with opposite phases)  is gapless with dispersion $q^{1/2}$, the QCM (like the QLCA) predicts a finite frequency $\omega(0)$ -- the gap -- in the $q\to0$ limit.  The origin of the gap in QCM is easily traced to short-range correlations between electrons and holes in opposite layers -- the enhancement of the ``on-top" electron-hole pair distribution function $g_{eh}(0)$ playing the key role.   The gap is  present for arbitrarily small values of the coupling parameter $r_s$,  although its value tends to zero very rapidly as $r_s \to 0$.  This suggest that the system is {\it never} a Fermi liquid -- a conclusion that may be formally correct for electron-hole bilayers.

The gap in the antisymmetric density excitation spectrum of electron-electron bilayer was first predicted in the classical (non-degenerate) regime \cite{kalman-ee}  and was directly confirmed by classical molecular dynamic simulation.  No such direct method is available for studying the dynamics of degenerate electron-hole systems. A semi-analytic study  \cite{gs-longitudinal,gs-transverse}  within the framework of QLCA (equivalent to the present approach) was carried out for degenerate {\it electron-electron} bilayers,  but not for degenerate electron-hole bilayers. 

There is an important difference between classical and quantum (degenerate) systems. 
In the classical system the finite temperature effectively erases the most significant signature of the gap, namely the rigidity (or incompressibility) of the system with respect to actions that attempt to change the densities of electrons and holes in opposite directions. The classical fluctuation-dissipation theorem~\cite{GV}
directly ties the density-density response function, which controls the response to external potentials,  to the static structure factor.  The latter vanishes linearly with $q$ for the antisymmetric channel in the $q\to0$ limit \cite{golden-2003}.   This indeed indicates  a suppression of the density response for small $q$, but its origin is quite clear: it is the large electrostatic energy associated with adding electrons to one layer and removing holes from the other.  In a realistic experimental setup,  this charging energy will be neutralized by additional charges occurring on the gates, ultimately leaving us with a finite density response (that is to say, the so-called ``proper" density-density response function~\cite{GV}, which does not vanish for $q\to0$).

In contrast to this, the density-density response function of the gapped quantum  system vanishes as $q^2$ for $q \to 0$. The vanishing density response is not caused by the charging energy, rather it reflects the energy cost of breaking bound electron-hole pairs as explained above.  This kind of rigidity cannot be compensated by additional charges on the gates: it is present in the proper density-density response function itself. 

Experimentally,  rigidity would show up in measurements of the electron and hole densities as functions of applied gate voltages $V_e$ and $V_h$.  Starting from the symmetric situation $n_e=n_h$, the application of potentials $V_e$ and $V_h$, which tend to shift the electron and hole densities in opposite directions, will have no effect on the densities until the potentials exceed a threshold given by the gap~\cite{Zeng2020}.

This paper is organized as follows.

In Section~\ref{secII} and appendix~\ref{Appendix1}, we review the elastic approximation of QCM and derive the formulas expressing the collective mode frequencies in terms of the pair distribution function and the  kinetic energy  of the ground state.  We show that our formulas coincide with those of the QLCA.

In Section~\ref{secIII}, we express the collective mode frequencies in terms of real-space integrals over the pair correlation functions and provide a simple formula, Eq.~(\ref{GapFormula}), expressing the gap in terms of the electron-hole interaction potential and the pair distribution function.

In subsection~\ref{secIII-A} we present and discuss the the evaluation of the real-space integrals  with pair correlation functions obtained from QMC, emphasizing the appearance of a gap in the antisymmetric density (longitudinal current) and   transverse current channel.  

In subsection~\ref{secIII-B}, the same calculations are performed with pair correlation functions obtained, at much lower computational cost, from the solution of the BCS-like mean field theory.  The pair correlation functions obtained in this manner are similar to the ones obtained in QMC, when the latter are available, and allow to explore the behavior of the gap in a wider region of values of $r_s$ and $d$. 

In Section~\ref{secIV}, we discuss the antisymmetric density response function obtained from QCM with the antisymmetric density response function obtained from the BCS-like mean field theory. Both  response functions are found to vanish in the long-wavelength limit, implying incompressibility.  The relation between the QCM gap and the BCS gap is clarified.

In Section~\ref{secV}  we present a critical discussion of the elastic approximation vis-a-vis the so-called Bijl-Feynman approximation~\cite{Mahan81} which produces a spectrum consistent with the exact static structure factor but in violation of the third moment sum rule.  

Section ~\ref{sumup} contains a summary  of our results for the excitation spectrum in the elastic approximation  and our outlook on  theoretical work going beyond the elastic approximation.

 \section {Electron-hole bilayer - Model and elastic approximation for collective modes}\label{secII}
 
 We consider a homogeneous symmetric electron-hole bilayer with hamiltonian
 \ber
 H&=&\sum_{i,\alpha} \frac{p_{i\alpha}^2}{2m} +\frac{1}{2}\sum_{i\neq j}\sum_{\alpha}\phi_{\alpha\alpha}(|\rv_{i,\alpha} - \rv_{j,\alpha}|)\nn\\
 &+&\sum_{i,j}\sum_{\alpha,\beta\neq\alpha}\phi_{\alpha\beta}(|\rv_{i,\alpha} - \rv_{j,\beta}|)\,,
 \eer
 where the indices $i$ and $j$ run over the particles (either electrons or holes)  of which there are equal numbers distributed with areal density $n=n_e=n_h$, and the indices $\alpha,\beta$  take values in the set $(e,h)$ where $e$ stands for electrons and $h$ for holes. Thus $\rv_{i,e}$ and $\pv_{i,e}$ are the (two-dimensional) position and  momentum of the $i$-th electron and similarly for holes. The mass $m$ is the same for electrons and holes.  The electron-electron and hole-hole interactions are
 \be\phi_{ee}(r)  = \phi_{hh}(r) =  \frac{e^2}{r}\,,
 \ee
 and
 \be 
 \phi_{eh}(r)=\phi_{he}(r) = -\frac{e^2}{\sqrt{r^2+d^2}}\,,
 \ee
 where  $r$ is the distance between two electrons (holes) in the same layer,  $d$ is the distance between the layers, and 
 $\sqrt{r^2+d^2}$ is the distance between an electron and a hole in opposite layers. $e$ is the absolute value of the electron charge.  The strength of the Coulomb interaction is measured by the Wigner-Seitz parameter 
\be
 r_s = \frac{1}{\sqrt{n \pi}a},
 \ee 
 where $a=\hbar^2/(me^2)$ is the Bohr radius. 
 
 We refer the reader to Ref.~(\cite{Gao10}) for a detailed discussion of QCM and the elastic approximation.  Here we only use the final result, which is an equation of motion for the displacement field $\uv(\rv,t)$, related to the current density by $\jv(\rv,t)=n(\rv)\dot\uv(\rv,t)$ where $n(\rv)$ is the ground state density and $\dot\uv(\rv,t)$ is the time derivative of the displacement field, i.e., the velocity of the volume element. 
 In the absence of external fields, this equation of motion has the form (see Eq. (46) in Ref.~(\cite {Gao10}))
 \be\label{EOM}
  m n(\rv)\ddot \uv(\rv,t)=  -\frac{\delta E_2[\uv]}{\delta\uv(\rv,t)}\,,
  \ee
  where  $E_2[\uv]$ is a quadratic functional of the displacement field, obtained by expanding the energy (kinetic plus potential) to second order in the displacement field.  (We note in passing that the system is assumed to have no net spin polarization: each volume element contains equal numbers of up- and down-spin particles, and they all follow the displacement field $\uv$ regardless of spin orientation). 
  
  The expression on the right hand side of Eq.~(\ref{EOM}) is a generalized force which, as shown in Ref.~(\cite{Gao10}), can be expressed in terms of the one-particle density matrix and the pair distribution function of the ground state.   Eq.~(\ref{EOM}) defines a small-oscillation problem whose eigenfrequencies are the excitation energies. For an isotropic system, such as a uniform electron liquid, the longitudinal component of $\uv$ yields the density  excitation spectrum, and the transverse component yields the transverse current excitation spectrum.  These approximate spectra consist of discrete excitation frequencies, at variance with the exact spectra, which are continuous. Yet, the spectral moments (first moment for the current fluctuation spectrum, first and third moment for the density fluctuation spectrum) can be shown to be exact, provided, of course, the input ground state information is exact. 
  
In order to apply the QCM formalism to the electron-hole bilayer we first generalize the formalism of  Ref.~(\cite{Gao10}) to  a system with several components in dimension D. This is done  in  appendix~\ref{Appendix1}. We then apply the formalism to a symmetric electron-hole system in 2 dimensions.  This involves  two displacement fields $\uv_e(\rv,t)$ and $\uv_h(\rv,t)$ for electrons and holes respectively.  They are related tho the electron and hole current densities by $\jv_\alpha(r,t)=n\dot\uv_\alpha(\rv,t)$, where $\alpha=e$ or $h$, and $n$ is the uniform density of electrons or holes.    The Eq.~(\ref{EOM}) becomes a system of two coupled linear equations after the replacement $\uv\to \uv_\alpha$ and the recognition that the energy $E_2$ is  a quadratic functional of both $\uv_e$ and $\uv_h$.
  
We take advantage of the translational invariance of the system (in the plane of the layers) by introducing the Fourier transform of the displacement fields, $\tilde \uv_\alpha(\qv,\omega)$, where $\qv$ is the wave vector and $\omega$ is the frequency.  The equation of motion takes the form
\be\label{EOM2}
-mn\omega^2\tilde\uv_\alpha =-\frac{\delta T_2[\tilde\uv_\alpha]}{\delta\tilde u_\alpha}-\frac{\delta W_2[\tilde\uv_e,\tilde \uv_h]}{\delta\tilde u_\alpha}\,,
\ee
where $T_2$ and $W_2$ are, respectively,  the kinetic and potential parts of the energy functional.
The functional derivatives on the right hand side of Eq.~(\ref{EOM2}) are readily obtained from the appropriate two-component generalization of Eqs. (53) and (58) of  Ref.~(\cite{Gao10}).  The formulas for the kinetic energy term are greatly simplified by dropping all the terms that contain the gradient of the ground state density.  The final expression is
\be
\frac{\delta T_2[\tilde\uv_\alpha]}{\delta\tilde \uv_\alpha} = nt(n)\left[2\qv(\qv\cdot\tilde \uv_\alpha)+q^2\tilde\uv_\alpha\right]+\frac{n\hbar^2q^2}{4m}\qv(\qv\cdot\tilde \uv_\alpha)\,,
\ee
with $\alpha=e,h$, where $nt(n)$ is the kinetic energy per unit area of the interacting electron-hole system -- a quantity well known from QMC calculations.   Notice that this ``kinetic force" does not couple the displacement fields of different species. 

\begin{widetext}
The calculation of the potential energy term is more complex but the final result is quite simple:
\be\label{GeneralizedForce}
\frac{\delta W_2[\tilde\uv_e,\tilde \uv_h]}{\delta\tilde \uv_\alpha} = \sum_{\beta,\nu} \left\{-[K_{\alpha\beta}(\qv={\bf 0})]_{\mu\nu}\tilde u_{\alpha\nu}(\qv)+[K_{\alpha\beta}(\qv)]_{\mu\nu}\tilde u_{\beta\nu}(\qv)\right\}\,, 
\ee
where  $\alpha$ and $\beta$ take values $e$ or $h$, and $\mu$ and $\nu$ are cartesian indices.    The all-important kernel $[K_{\alpha\beta}(\qv)]_{\mu\nu}$ is calculated from the structure factors of the ground state, $S_{\alpha\beta}(q)$, as follows
 \be\label{InteractionKernel}
 [K_{\alpha\beta}(\qv)]_{\mu\nu}=n\int \frac{d\qv'}{(2\pi)^2}\left[S_{\alpha\beta}(|\qv-\qv'|)-\delta_{\alpha\beta}\right]\tilde \phi_{\alpha\beta}(q')q'_{\mu}q'_{\nu} +n^2q_{\mu}q_{\nu}\tilde\phi_{\alpha\beta}(q),
 \ee
 where
\be
S_{\alpha\beta}(\qv)=\frac{1}{N}\sum_{n,n'}\left\langle e^{i\qv\cdot(\rv_{n,\alpha}-\rv_{n'\beta})}\right\rangle -N\delta_{{\bf q},0},
\ee
$\langle...\rangle$ denotes the ground state average and  $\tilde\phi_{ee}(q)=\tilde\phi_{hh}(q)=\frac{2\pi e^2}{q}$,  $\tilde\phi_{eh}(q)=\tilde\phi_{he}(q)=-\frac{2\pi e^2}{q}e^{-qd}$..
Notice that the structure factors  are expressible in terms of the Fourier transforms of the pair distribution functions for species $\alpha$ and $\beta$~\cite{GV}: we will make use of this in the next section.

The symmetry of the problem ($S_{ee}=S_{hh}$, $S_{eh}=S_{he}$)  allows us to decouple the equations of motion into symmetric/antisymmetric channels denoted by $+$ and $-$ respectively,  defined as follows
\be
\tilde u_{\pm}(\qv,\omega)=\tilde u_{e}(\qv,\omega)\pm \tilde u_{h}(\qv,\omega)\,.
\ee
 Furthermore, isotropy allows us to decouple the longitudinal channel ($\tilde \uv \parallel \qv$), denoted by $L$,  from the transverse channel
 ($\tilde \uv \perp \qv$), denoted by $T$.  Thus we arrive at the following explicit formulas for the frequencies of the longitudinal collective modes
 \ber\label{LP}
 \omega_{L+}^2(\qv)&=&\frac{2\pi n e^2 q}{m}\left(1-e^{-qd}\right)+ q^2\left[\frac{3t(n)}{m}+\frac{\hbar^2q^2}{4m^2}\right]+\frac{1}{m}\int \frac{d\qv'}{(2\pi)^2}\left[S_{ee}(|\qv-\qv'|)-S_{ee}(q')\right]\tilde \phi_{ee}(q')(\qv'\cdot\hat \qv)^2\nn\\ 
 &+&\frac{1}{m}\int \frac{d\qv'}{(2\pi)^2}\left[S_{eh}(|\qv-\qv'|)-S_{eh}(q')\right]\tilde \phi_{eh}(q')(\qv'\cdot\hat \qv)^2\,,
 \eer
 and
 \ber\label{LM}
 \omega_{L-}^2(\qv)&=&\frac{2\pi n e^2 q}{m}\left(1+e^{-qd}\right)+ q^2\left[\frac{3t(n)}{m}+\frac{\hbar^2q^2}{4m^2}\right]+\frac{1}{m}\int \frac{d\qv'}{(2\pi)^2}\left[S_{ee}(|\qv-\qv'|)-S_{ee}(q')\right]\tilde \phi_{ee}(q')(\qv'\cdot\hat \qv)^2\nn\\ 
 &-&\frac{1}{m}\int \frac{d\qv'}{(2\pi)^2}\left[S_{eh}(|\qv-\qv'|)+S_{eh}(q')\right]\tilde \phi_{eh}(q')(\qv'\cdot\hat \qv)^2\,,
 \eer
where $\hat \qv$ is the unit vector along $\qv$.   Similarly, for the transverse collective modes we may set ${\bf q}=q\hat{x}$ to obtain
\ber\label{TP}
 \omega_{T+}^2(\qv)&=& q^2\frac{t(n)}{m}+\frac{1}{m}\int \frac{d\qv'}{(2\pi)^2}\left[S_{ee}(|\qv-\qv'|)-S_{ee}(q')\right]\tilde \phi_{ee}(q')(q'_y)^2\nn\\ 
 &+&\frac{1}{m}\int \frac{d\qv'}{(2\pi)^2}\left[S_{eh}(|\qv-\qv'|)-S_{eh}(q')\right]\tilde \phi_{eh}(q')(q'_y)^2\,,
 \eer
 and
 \ber\label{TM}
 \omega_{T-}^2(\qv)&=&q^2\frac{t(n)}{m}+\frac{1}{m}\int \frac{d\qv'}{(2\pi)^2}\left[S_{ee}(|\qv-\qv'|)-S_{ee}(q')\right]\tilde \phi_{ee}(q')(q'_y)^2\nn\\ 
 &-&\frac{1}{m}\int \frac{d\qv'}{(2\pi)^2}\left[S_{eh}(|\qv-\qv'|)+S_{eh}(q')\right]\tilde \phi_{eh}(q')(q'_y)^2\,.
 \eer
   \end{widetext}
 The main qualitative features of the spectrum are immediately visible in these formulas.    The symmetric channel spectrum is gapless, because the expressions  $\left[S_{ee}(|\qv-\qv'|)-S_{ee}(q')\right]$ and  $\left[S_{eh}(|\qv-\qv'|)-S_{eh}(q')\right]$ in the integrals of Eqs.~(\ref{LP}) and (\ref{TP}) vanish for $q \to 0$.
 In contrast,  the antisymmetric channel spectrum, in which electrons and holes oscillate with opposite phases in the two layers, is gapped because the expression $\left[S_{eh}(|\qv-\qv'|)+S_{eh}(q')\right]$ in the integrand of  Eqs.~(\ref{LM}) and (\ref{TM}) {\it does not vanish} for $q\to 0$.  The existence of short-range electron-hole correlation, described by the structure factor $S_{eh}(q)$ and the associated pair distribution function $g_{eh}(r)$ is fully responsible for the emergence of  the gap.

 \section {Real-space Implementation}\label{secIII}
We now calculate the frequency of the collective modes, Eqs.~(\ref{LP}-\ref{TM}), using real-space  pair-correlation functions $g_{eh}(r)$ and $g_{ee}(r)$ \cite{tramonto}.    For example, the integral
\be\label{MainIntegral}
I^{\pm}_{\alpha\beta}(q)=\int \frac{d\qv'}{(2\pi)^2}\left[S_{\alpha\beta}(|\qv-\qv'|)\pm S_{\alpha\beta}(q')\right]\tilde \phi_{\alpha\beta}(q')(\qv'\cdot\hat \qv)^2
\ee  
can be rewritten as
\begin{widetext} 
\be\label{RealSpaceIntegral}
I^{\pm}_{\alpha\beta}(q)=2\pi n\int_0^\infty dr r h_{\alpha\beta}(r)\left\{\frac{1}{2}\left[J_0(qr)-J_2(qr)\pm 1\right]\left[\frac{\phi_{\alpha\beta}^{\prime}(r)}{r}-\phi_{\alpha\beta}^{\prime\prime}(r) \right]-\left[J_0(qr)\pm 1\right]\frac{\phi_{\alpha\beta}^{\prime}(r)}{r}\right\}\,,
\ee
\end{widetext} 
where $h_{\alpha\beta}(r)=g_{\alpha\beta}(r)-1$ and $J_n(qr)$ are Bessel functions of order $n$.  

Notice that the $\pm$ signs in Eq.~(\ref{RealSpaceIntegral}) make the all-important difference between the gapless  spectrum in the symmetric channel and the gapped one in the antisymmetric channel. The square of the gap is given by 
\be\label{GapFormula} 
\omega^2(0)=-\frac{1}{m}I^+_{e,h}(0)=\frac{n}{m}\int d{\bf r}\, h_{e,h}(r)\nabla^2\phi_{e,h}(r),
\ee
which (in Rydberg$^2$) is given by the simple formula  
\be\label{GapFormula2}
\omega^2(0)=\frac{8}{r_s^2}\int _0^\infty  dr h_{eh}(r)  r\frac{2d^2-r^2}{(r^2+d^2)^{5/2}}\,,
\ee
where  $r, d$ are in atomic units.
This has the same value for the longitudinal and the transverse mode.

\subsection{Calculation with QMC pair correlation functions}\label{secIII-A}
\begin{figure*}
\centering
\includegraphics[width=\textwidth]{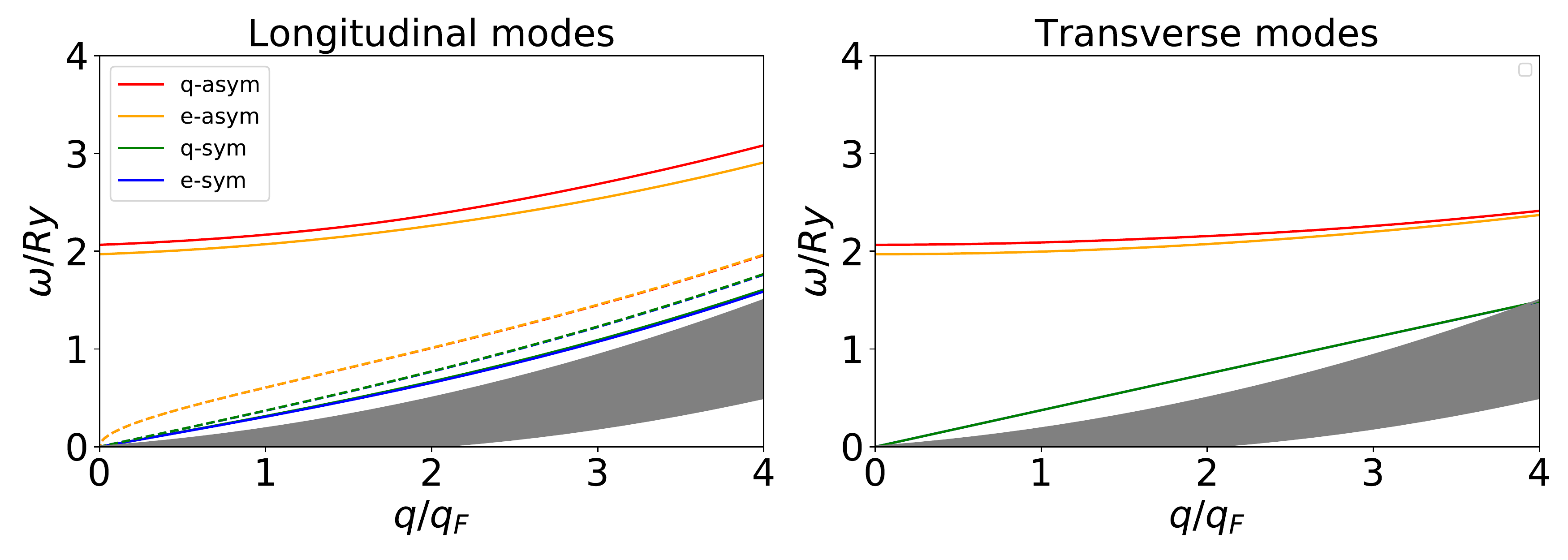}
\caption{Interlayer distance $d=0.3$ a.u. Left panel: longitudinal modes $\omega_{L-}(q)$, $\omega_{L+}(q)$ ; right panel: transverse modes $\omega_{T-}(q)$, $\omega_{T+}(q)$.  The non-interacting single-particle spectrum is shown for reference as a shaded region. Solid lines: dispersion of the  antisymmetric  mode   in the quadriexcitonic  (q) phase (red lines) and  in the excitonic (e) phase  (orange lines); dispersion of  the symmetric   mode  in the quadriexcitonic phase (green lines) and in the excitonic phase (blue lines). Dashed lines  with the same color coding  present  the dispersion of the modes calculated without   the structure factor terms in eqs. \eqref{LP},\eqref{LM},\eqref{TP},\eqref{TM}.}
\label{fig:f2}
\end{figure*}

\begin{figure*}
\centering
\includegraphics[width=\textwidth]{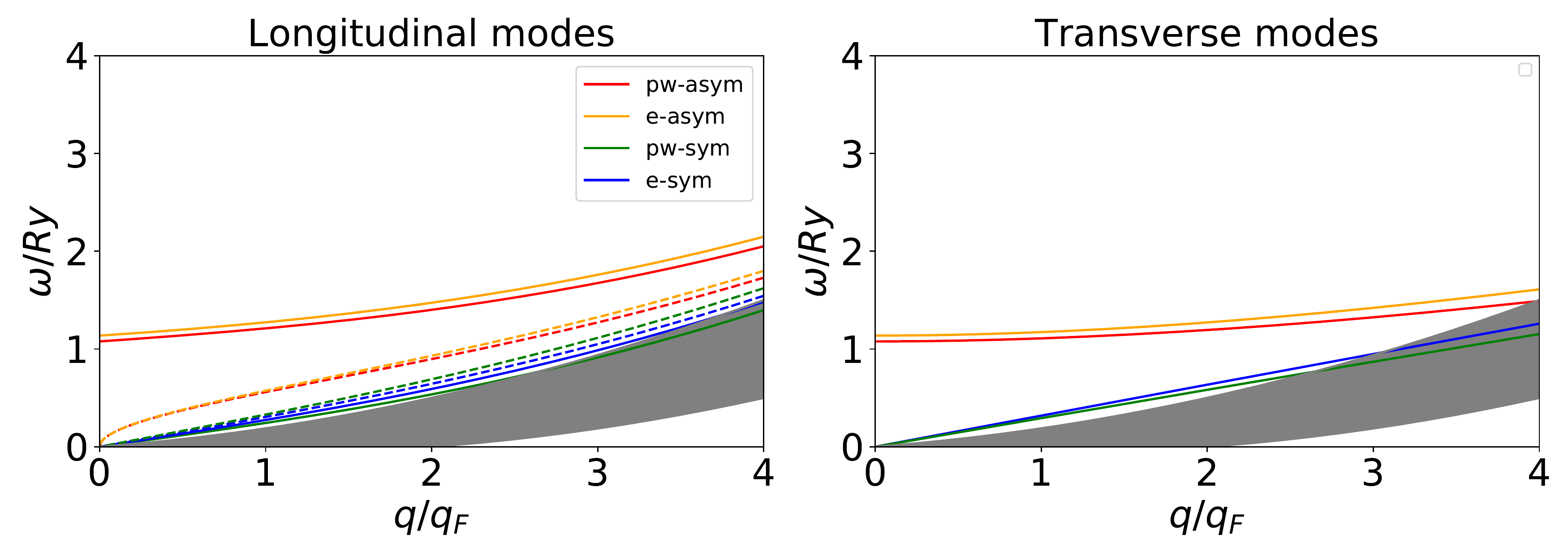}
\caption{Interlayer distance $d=0.5$ a.u. Left panel: longitudinal modes $\omega_{L-}(q)$, $\omega_{L+}(q)$ ; right panel: transverse modes $\omega_{T-}(q)$, $\omega_{T+}(q)$.  The non-interacting single-particle spectrum is shown for reference as a shaded region.
Solid lines: dispersion of the  antisymmetric  mode   in the excitonic  (e) phase (orange lines) and  in the plasma (pw) phase  (red lines); dispersion of  the symmetric   mode  in the excitonic phase (blue lines) and in the plasma phase (green lines). Dashed lines  with the same color coding  present  the dispersion of the modes calculated without   the structure factor terms in eqs. \eqref{LP},\eqref{LM},\eqref{TP},\eqref{TM}.}
\label{fig:f3}
\end{figure*}

\begin{figure*}
\centering
\includegraphics[width=\textwidth]{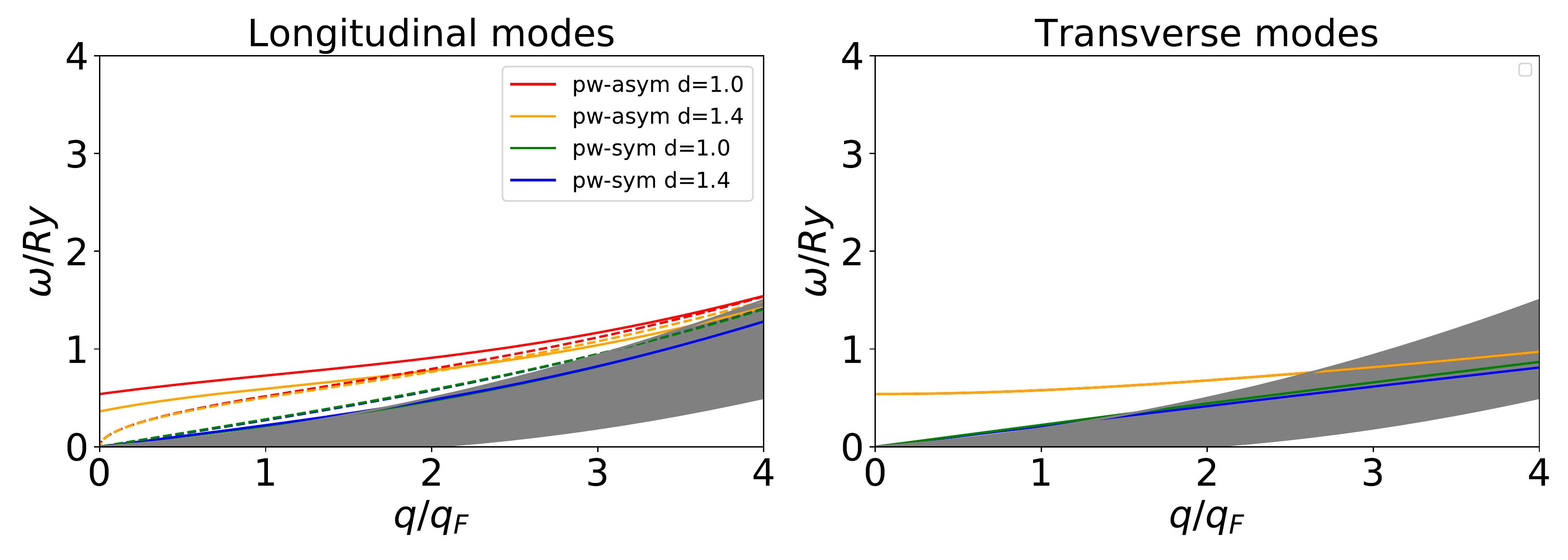}
\caption{Interlayer distances $d=1.0$ a.u. and $d=1.4$ a.u, plasma phase (pw). Left panel: longitudinal modes $\omega_{L-}(q)$, $\omega_{L+}(q)$ ; right panel: transverse modes $\omega_{T-}(q)$, $\omega_{T+}(q)$.  The non-interacting single-particle spectrum is shown for reference as a shaded region.
Solid lines: dispersion of the  antisymmetric  mode   for interlayer distance $d=1.0$ a.u. (red lines)  and  $d=1.4$ a.u. (orange lines); dispersion of  the symmetric   mode for interlayer distance $d=1.0$ a.u. (green  lines)  and  $d=1.4$ a.u. (blue lines). Dashed lines  with the same color coding  present  the dispersion of the modes calculated without   the structure factor terms in eqs. \eqref{LP},\eqref{LM},\eqref{TP},\eqref{TM}.}
\label{fig:f4}
\end{figure*}

In Figs~\ref{fig:f2}, \ref{fig:f3}, and \ref{fig:f4}  we plot the dispersion of the symmetric and antisymmetric modes, longitudinal (left panels) and transverse (right panels) obtained by evaluating the real-space integrals ~(\ref{RealSpaceIntegral}) with QMC pair correlation functions.  Results are presented for  $r_s=4$ and four different values of the interlayer distance $d=0.3$ a.u.,   $d=0.5$ a.u.,  $d=1.0$ a.u.,  $d=1.4$ a.u.   Because the QMC calculations assumed two equivalent valleys and two equivalent spin orientations in each layer, the Fermi wave vector $q_F$, which we use as the wave vector unit in these figures, is related to the density by $q_F=\sqrt{\pi n}$.

It is evident that the asymmetric modes display a finite gap for $q\rightarrow 0$. Such a gap is largest at the smallest distance considered ($d=0.3$) and decreases for larger distances. 


\begin{figure}[ht]
\centering
\includegraphics[width=0.5\textwidth]{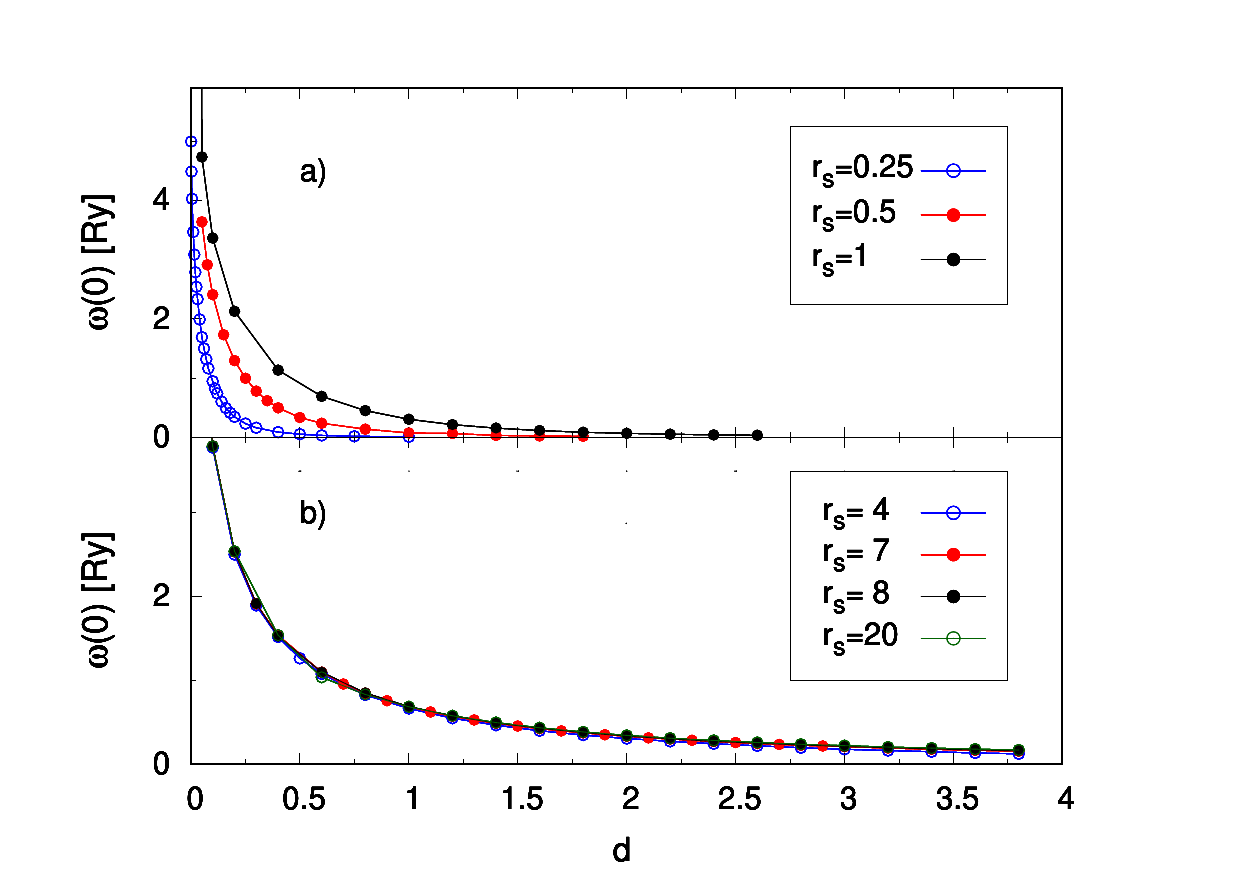}
\caption{Plot of $\omega(0)$ vs $d$ for different values of $r_s$.  $\omega(0)$ is calculated from Eq.~(\ref{GapFormula2}) and the pair correlation function is obtained from the self-consistent solution of the BCS mean field theory, according to Eq.~(\ref {geh(r)-BCS}), as explained in the text.}
\label{fig:f5}
\end{figure}

\subsection {Calculation with BCS pair correlation functions}\label{secIII-B}
The QMC calculation of the pair correlation functions is time-consuming and the results are available only for a few values of $r_s$ and $d$. We achieve much greater flexibility by resorting to the BCS-like mean field theory of Ref.~   \cite{Littlewood1995,Littlewood1996}.  The crucial pair correlation function $h_{eh}(r)$ in this approach is given by 

\be\label{geh(r)-BCS}
h^{BCS}_{eh}(r)=\left\vert\frac{1}{n}\int_0^\infty \frac{dk k}{2\pi}\frac{\Delta (k)}{E(k)} J_0(kr)\right\vert^2,
\ee
where the BCS gap function $\Delta(k)$ and the BCS quasiparticle energy $E(k)=\sqrt{\xi(k)^2+\Delta(k)^2}$ are obtained from the self-consistent solution of the mean field equations~\cite{Littlewood1995,Littlewood1996} where the 
$k$-dependence of the gap, as well as the in-plane interaction, are taken into account. 

$h_{eh}(r)$ calculated in this manner is in quite good agreement with the QMC result in the cases  in which the latter is available.  More importantly, using $h^{BCS}_{eh}(r)$ in Eq.~(\ref{GapFormula2}) we can calculate $\omega(0)$ for a broad range of values of $r_s$ and $d$.  The results of the calculation are plotted in Fig. \ref{fig:f5}.
A remarkable feature of these results is that in the low-density limit ($r_s \to \infty$) $\omega(0)$ approaches a finite limit, independent of $r_s$.  This is physically expected since in this limit the system reduces to a collection of well-separated bound electron-hole pairs and the pair correlation function becomes
\be
\label{eq:heh_exc}
h_{eh}(r)= n^{-1}|\psi_{eh}(r)|^2\,,
\ee
where $\psi_{eh}(r)$ is the wave function of the bound state.  In the opposite limit of high density ($r_s \to 0$) and finite $d$ the BCS gap goes to zero and so does $\omega(0)$.

Another interesting feature of the result is the logarithmic divergence of $\omega(0)$ in the limit of vanishing interlayer separation ($d\to0$).  Mathematically, this arises from the two-dimensional integration of the Coulomb interaction, which diverges at $r=0$.  This logarithmic divergence is likely to be an artifact of the elastic approximation.  This will become more evident in the next section, where we relate $\omega(0)$ to the long-wavelength behavior of the antisymmetric density response function and hence to the incompressibility. 

\section{Antisymmetric density response and incompressibility}\label{secIV}
 
 In this section we focus on the longitudinal spectrum in the antisymmetric channel.  Unless otherwise indicated $\omega_q$ will be a shorthand for $\omega_{L-}(q)$.  Similarly $\chi(q,\omega)$ will be a shorthand for $\chi_{L-}(q,\omega)\equiv \chi_{ee}(q,\omega) -\chi_{eh}(q,\omega)$ and $S(q)$ will be a shorthand for $S_{L-}(q)\equiv S_{ee}(q)-S_{eh}(q)$.
 
 It is well-known~\cite{Gao10} that the elastic approximation for a one component system satisfies the so-called ``third moment" sum rule for the density fluctuation  spectrum. In our two-component system  the sum rule reads
\be\label{third-msum}
 -\frac{2}{\pi}\int_0^\infty d\omega \omega^3 \Im m\chi(q,\omega) = M(q)q^2,
 \ee
 where $M(q)$ (standing for $M_{L-}(q)$, as explained above)  is expressed in terms of the exact kinetic energy and  structure factors as ~\cite{GV,nota1} \begin{widetext}
 \ber\label{LM3}
M(q)&=&\frac{n}{m}\left\{\frac{2\pi n e^2 q}{m}\left(1+e^{-qd}\right)+ q^2\left[\frac{3t(n)}{m}+\frac{\hbar^2q^2}{4m^2}\right]+\frac{1}{m}\int \frac{d\qv'}{(2\pi)^2}\left[S_{ee}(|\qv-\qv'|)-S_{ee}(q')\right]\tilde \phi_{ee}(q')(\qv'\cdot\hat \qv)^2\right.\nn\\ 
 &-&\left.\frac{1}{m}\int \frac{d\qv'}{(2\pi)^2}\left[S_{eh}(|\qv-\qv'|)+S_{eh}(q')\right]\tilde \phi_{eh}(q')(\qv'\cdot\hat \qv)^2\right\}\,.
 \eer
\end{widetext}
On the other hand, a simple calculation starting from the equation of motion for the antisymmetric longitudinal mode, in the presence of  external potentials acting on  particles and identifying  $-i n {\bf q}\cdot{\bf u}_{\gamma}$  with  the density change $\delta n_{\gamma}$, allows the calculation of the response function in the QCM approximation as  
\be\label{SpectralFunctionEA}
 -\frac{1}{\pi}\Im m\chi_{QCM}(q,\omega) =  \frac{ n q^2}{2m\omega_q}[\delta(\omega-\omega_q) -  \delta(\omega+\omega_q)]\,.
 \ee 
 It is immediately verified that the QCM harmonic response satisfies  both the $f$-sum rule~\cite{GV}
  \be
  -\frac{2}{\pi}\int_0^\infty d\omega \omega \Im m\chi_{QCM}(q,\omega) = \frac{nq^2}{m},
 \ee
and the third moment sum rule
\be
-\frac{2}{\pi}\int_0^\infty d\omega \omega^3 \Im m\chi_{QCM}(q,\omega) =\frac{nq^2}{m}\omega_q^2=M(q)q^2,
\ee
as from eqs. (\ref{LM}) and (\ref{third-msum})
 \be\label{omegaqEA}
  \omega_q = \sqrt{\frac{m M(q)}{n}}.
  \ee
 Notice that $M(q)$ tends to a finite limit for $q\to 0$, as discussed above.  This is a unique feature of the antisymmetric density channel: in the symmetric density channel, the corresponding quantity $M_{L+}(q)$ is known to vanish as $q^2$ for $q\to 0$.
 
  These reassuring results help us  understand why our formulas for the collective mode frequencies coincide with those of Kalman et al., obtained from the QLCA. Both theories collapse the spectrum onto a single collective mode which satisfies the third and first moment sum rules: these constraints are strong enough to uniquely determine the frequencies.

Let us now consider the static density response, which is given by the dispersion relation
\be \label{static-chi}
\chi(q,0)=\frac{2}{\pi}\int_0^\infty d\omega \frac{\Im m\chi (q,\omega)}{\omega}\,.
\ee
(notice that this can be considered a ``negative-first-moment sum rule"). 
Use of  Eq. (\ref{SpectralFunctionEA}) yields  the  static response in QCM:
 \be\label{chiq0}
 \chi_{QCM}(q,0)= - \frac{nq^2}{m\omega_q^2}\,.
 \ee
The existence of a finite gap ($\omega_q\to \omega_0\equiv\omega(0)>0$) for $q\to0$ immediately implies that $\chi_{QCM}(q,0)$ vanishes as $q^2$.    We emphasize that the vanishing of $\chi_{QCM}(q,0)$ is stronger than what would be expected from purely electrostatic considerations, e.g., from the random phase approximation, if the compressibility remained finite.  
Indeed, in the $q\to 0$ limit the static response is related to the compressibility $K_-$ (in the antisymmetric density channel) by the relation
\be\label{StaticChi}
 \chi(q,0) \stackrel {q\to 0} {\rightarrow} -\frac{1}{v_{q-}+\frac{1}{n^2K_-}}\,, 
 \ee
 where $v_{q-}=\frac{2\pi e^2}{q}(1+e^{-qd})$ and
 \be
 \frac{1}{n^2 K_-}=\left.2\frac{\partial^2 \epsilon(n,\delta n_-)}{(\partial \delta n_-)^2}\right\vert_{n_e=n_h=n}\,,
 \ee
 is twice the second derivative of the energy density with respect to the imbalance density $\delta n_-=n_e-n_h$ evaluated at the charge neutrality point. (Notice that  $\epsilon$ in this formula is the energy of a charge-neutral system, that is to say, we assume that the charge imbalance associated with $\delta n_-$ is neutralized by  compensating background charges at zero energy cost.) 
 Now if $K_-$ were finite, then $\chi(q,0)$ would vanish as $q$, due to the divergence of $v_{q-}$ in the denominator of Eq.~(\ref{StaticChi}). But we have seen that  $\chi(q,0)$ vanishes as $q^2$:  this can be reconciled only if $K_-$ is zero. More pointedly, we could introduce a $q$-dependent compressibility $K_-(q)$, whose inverse is the second derivative of the energy with respect to the amplitude of a density fluctuation of wave vector $q$. This $q$-dependent compressibility, replacing $K_-$ in Eq.~(\ref{StaticChi}) would  vanish as $q^2$ in the $q \to 0$ limit.

 The QCM prediction of the existence of a gap $\omega(0)$  in the uniform (non-BCS) plasma phase of the electron-hole liquid is quite surprising: in this phase one would expect to find a gapless plasmon, dispersing as $q^{1/2}$.  On the other hand, the existence of a gap is completely expected in the paired BCS or excitonic phase,  because the formation of bound electron-hole pairs prevents long-range charge separation.
In  Section \ref{secIII}, we  used the BCS model  as a practical tool to calculate pair correlation functions to be fed into the QCM machinery. Now we proceed to a more direct comparison between the physical predictions of QCM and  BCS-like mean field theory. In particular, we compare the antisymmetric density response obtained in QCM (see Eq.~(\ref{chiq0})) with the same response calculated within the BCS-like mean field theory.

 We start from the observation that in the BCS-like theory 
 \cite{Littlewood1995,Littlewood1996},  the density response function in the antisymmetric channel is given by
\be
\label{chiBCS}
\chi(q,0)=-q^2\frac{g_c}{32\pi}\int_0^{\infty} dk k\,  \frac{[\xi(k)\Delta'(k)-\xi'(k)\Delta(k)]^2}{[\Delta(k)^2+\xi(k)^2]^{5/2}},
\ee
where $g_c$ is the number of fermionic components per layer (here $g_c=4$) and the quantities $\xi(k)$ and $\Delta(k)$ here are per particle, i.e., a half of those in Ref. \onlinecite{Littlewood1995}, which are per exciton (electron-hole pair). The prime denotes the derivative of a function with respect to its own argument. 

The static response of the excitonic state assumes a particularly transparent form in the low-density limit, whereby to leading order in the density $n$ it becomes 
\be
\label{lowden}
\chi(q,0)=-q^2\frac{g_cn}{32\pi|\mu|}\int_0^{\infty} dk k\,  \frac{[\tilde\psi_{eh}'(k)]^2}{1+(\xi k)^2}.
\ee
In the above formulas  $|\mu|$ and $\tilde\psi_{eh}(k)$ are,   respectively,    half the binding energy  of the isolated exciton  and  the Fourier transform of its normalized wavefunction; also,  $\xi^2=\hbar^2/(2 m |\mu|)$.
For zero interlayer distance the wavefunction of the isolated exciton ($\mu=-\hbar^2/(2ma_B^2$)) is  known in closed form, $\tilde\psi_{eh}(k)=2\sqrt{2 \pi}a_B/[1+(ka_B)^2]^{3/2}$.  The integral in Eq.(\ref{lowden}) can be readily performed and by comparison with Eq. (\ref{chiq0}) one gets $\hbar\omega(0)=2\sqrt{(20/9)}Ry$.

In the opposite  limit of high density (small $r_s$)  there is no known simplification of Eq. (\ref{chiBCS}). However we have found that even at $d=0$ the numerical solution with a finite gap function to the BCS  equations \cite{Littlewood1995}  is  lost  for  $r_s\lesssim 0.04$, implying that the system turns to the normal state. Thus  at least at the RPA level the  static response becomes  linear in $q$, implying the disappearance of the gap, i.e.,  $\omega(0)=0$. 

\begin{figure}
\centering
\includegraphics[width=8cm]{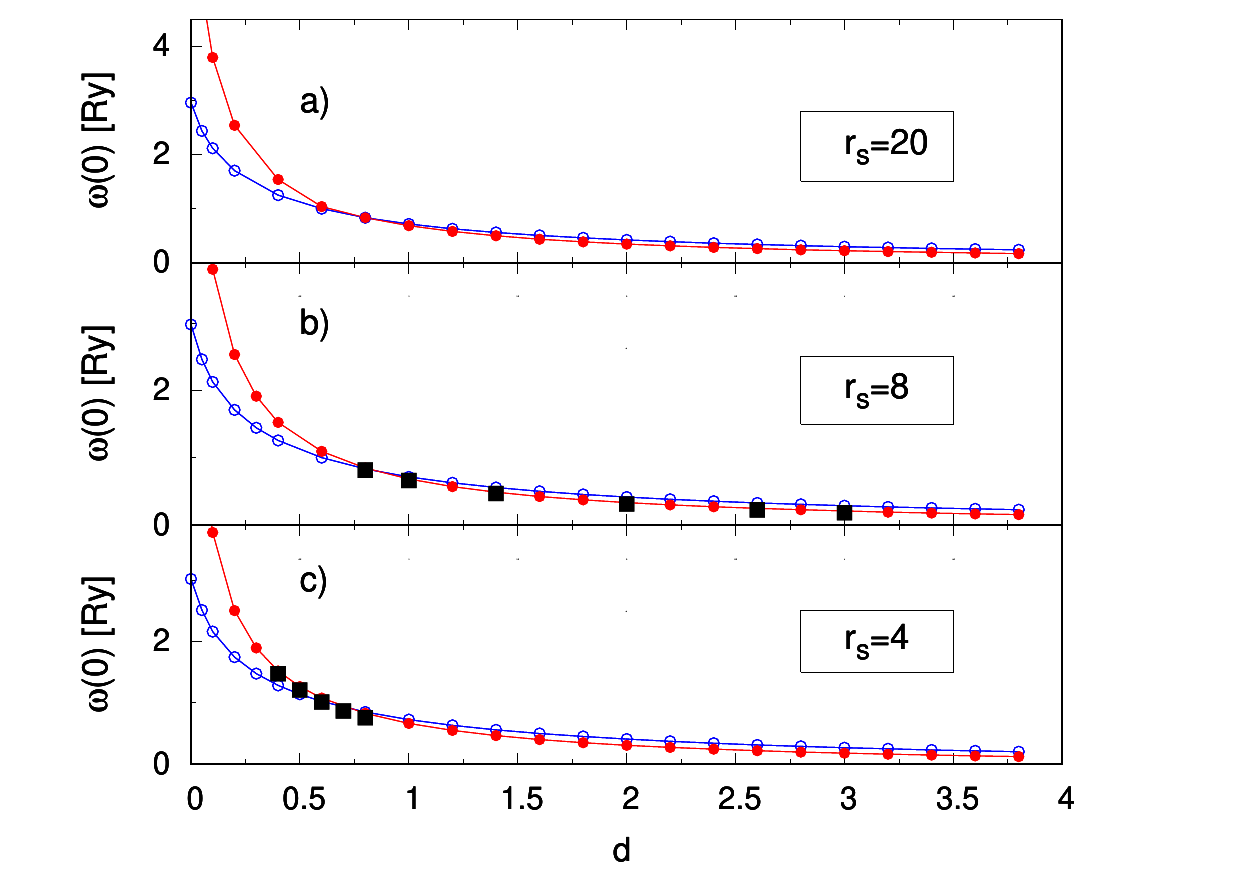}
\caption{Plot of $\omega(0)$ as a function of distance $d$ for $r_s=20,8$ and $4$, respectively in 
panels $a), b)$ and $c)$. The solid red dots are calculated from
Eq.~(\ref{GapFormula2}) using the pair correlation function obtained from the 
self-consistent solution of the BCS mean field theory, according to Eq.~(\ref {geh(r)-BCS}).  
The solid black squares are obtained using the pair correlation functions from QMC simulations \cite{tramonto}.
The open blue dots are the values of $\omega(0)$ obtained from the direct comparison of Eqs.~(\ref{chiBCS}) and (\ref{chiq0})}
\label{fig:f6}
\end{figure} 

\begin{figure}
\centering
\includegraphics[width=9 cm]{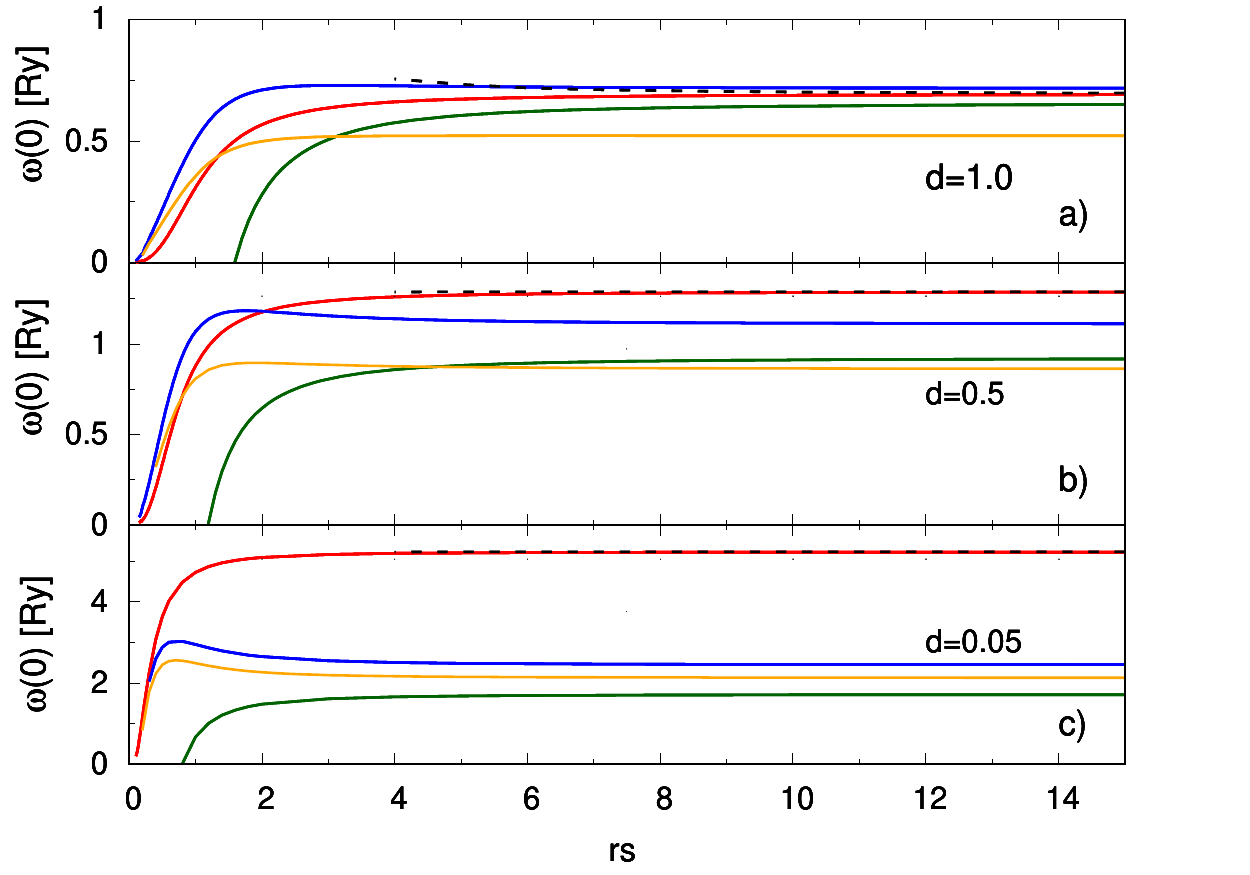}
\caption{ Plot of $\omega(0)$ as a function of $r_s$ for various values of the distance $d$. Solid red and blue lines are, respectively, for $\omega_{QCM}(0)$ and $\omega_{\chi}(0)$.  The solid orange line is for $\omega_{BF}(0)$, which is the $q\to0$ limit of the Bijl-Feynman frequency, Eq.~(\ref{BFDispersion}), calculated with the BCS structure factor of Eq.~(\ref{SqBCS}), as explained in Section~\ref{secV}.  The dashed black line is  $\omega_{QCM}(0)$ for a system made of isolated excitons, with the $h_{eh}(r)$ from Eq.~(\ref{eq:heh_exc}). 
The solid 
dark-green line is $-2 \mu$.
Panels $a) , b)$ and $c)$ show data 
for $d=1,0.5$ and $d=0.05$ respectively.}
\label{fig:f7}
\end{figure}

 In  Figs.~\ref{fig:f6} and~\ref{fig:f7} we show the values of $\omega(0)$   obtained from the direct comparison of Eqs.~(\ref{chiBCS}) and (\ref{chiq0}) (blue curves) versus the values of $\omega(0)$  obtained from Eq.~(\ref{GapFormula2}) together with the pair correlation function given by Eq.~(\ref{geh(r)-BCS}) (red curves).  In the following, the gap obtained from the BCS response will be referred to as $\omega_{\chi}(0)$ and the one obtained from the quantum continuum mechanics will be referred to as $\omega_{QCM}(0)$. The agreement between the two sets of values is generally quite good, particularly for large $r_s$ and $d$. However, for small $d$, the QCM gap (red curve) is much too large in comparison with the one from the BCS response: this reflects the existence of an artificial logarithmic divergence of $\omega_{QCM}(0)$ for $d \to 0$, as was pointed out at the end of Section~\ref{secIII}.  Interestingly, there are also regions of parameter space (small $r_s$, large $d$) in which the QCM gap is {\it smaller} than the one from the BCS response: however, the difference between the two estimations of the gap is quite small in these regions.
 
Fig.~\ref{fig:f7} shows the existence of a tight connection between $\omega(0)$ and the exciton binding energy in the low density limit and for large $d$.  This is illustrated by the solid green line which shows  $-2\mu$ vs $r_s$, with $\mu$ the chemical potential calculated within BCS theory.  As discussed in the introduction,  the binding energy of an electron-hole pair is given by $\epsilon_e+\epsilon_h-2\mu$ where $\epsilon_e,\epsilon_h$ are the energies of free (unbound) electrons and holes respectively.  In the low-density limit $\epsilon_e$ and $\epsilon_h$ are the energies of electrons and holes at the bottom of the respective bands, which we take to be at zero energy. Thus $-2\mu$ is the binding energy of the pair in this limit: indeed, we see that it approaches $\omega(0)$  for large $r_s$ and $d$.    This is important because the exciton binding energy is a quantity that can be determined experimentally, by measuring the variations of the densities of electrons and holes as we  apply potentials $V_e$ and $V_h$ which push them in opposite directions (see  Fig. 1). 
Incompressibility means that the system will resist the push and remain balanced ($n_e=n_h$) in a range of potentials of order $\epsilon_b$ due to the rigidity associated with electron-hole pairing, as explained in the introduction.  See Ref.~\onlinecite{Zeng2020} for details.

 \section{Discussion}\label{secV}
The Quantum Continuum Mechanics is a formally exact theory of quantum dynamics\cite{Gao10}, but the elastic approximation, on which the present work is based, makes the questionable assumption that the time-dependent wave function is, at every instant of time $t$, obtained  by applying a deformation with displacement field $\uv(\rv,t)$ to the ground state wave function.  Under this assumption the excitation spectrum is reduced to a set of sharp normal modes, which describe collective density and current oscillations. This collective description of the dynamics is expected to be qualitatively correct in strongly correlated systems, where the single-particle degrees of freedom are effectively suppressed as the individual particles are ``enslaved" to  collective modes.  In this section we briefly discuss some issues that arise in connection with the approximate character of the theory.

  One drawback of the elastic approximation becomes evident when we consider the fluctuation-dissipation theorem, which actually takes the form of a zeroth-moment sum rule or ``$S$-sum rule" as follows:
  \be
 -\frac{\hbar}{\pi n}\int_0^\infty d\omega \Im m\chi(q,\omega) = S(q)\,.
 \ee
As in the previous section $\Im m\chi(q,\omega)$ and $S(q)$  denote the spectral density and the structure factor in the {\it antisymmetric} density-fluctuation channel.  In general the  QCM formula for the spectral function in the elastic approximation fails to satisfy this sum rule, even as it satisfies the third moment sum rule exactly. For example, in the high-density degenerate plasma phase, where the random phase approximation is valid,  we know that the exact $S(q)$ must tend to zero as $q^{3/2}$ for $q \to 0$.
But even in that case the elastic approximation predicts a spectral gap, and hence $S(q) \propto q^2$ for $q\to 0$, as we can easily verify from our expression~(\ref{SpectralFunctionEA}).

An approximation that satisfies the $S$-sum rule (but violates the third-moment sum rule) is known as Bijl-Feynman (BF) approximation~\cite{Mahan81} and yields the following formula for the spectral function
\be\label{SpectralFunctionBF}
 -\frac{1}{\pi}\Im m\chi_{BF}(q,\omega) =  \frac{n}{\hbar} S(q)[\delta(\omega-\omega_{BF}(q))-\delta(\omega+\omega_{BF}(q))\,,
 \ee
 with
 \be\label{BFDispersion}
 \omega_{BF}(q)\equiv \frac{\hbar q^2}{2mS(q)}\, ,
 \ee
to satisfy the f-sum rule.  In the  uniform plasma phase (without BCS pairing) $S(q)$  vanishes as $q^{3/2}$ for $q\to0$ -- a property that reflects the electrostatic energy of long-range charge density fluctuations.  Thus, the BF dispersion of the uniform plasma phase is necessarily gapless and goes as $q^{1/2}$ for $q\to0$ as expected for the classical plasmon in two dimensions.     Similarly, if we consider the static density-density response function $\chi(q,0)$,  obtained from Eq.~(\ref{static-chi}), we see that the BF approximation predicts $\chi_{BF}(q,0) \sim q \sim 1/v_{q-}$.  Comparison with Eq.~(\ref{StaticChi})  shows that the antisymmetric compressibility $K_-$ it may vanish at most as $q^{\alpha}$ with $0\leq \alpha < 1 $.

Compared to the Bijl-Feynman approximation, the elastic approximation of QCM has two major advantages.  First, it gives  a richer spectrum of collective modes, including both longitudinal and transverse excitations.  Second, it expresses the frequencies of collective modes in terms of real space integrals, which are dominated by the short-range part of the pair correlation functions (see, for example, Eqs.~(\ref{GapFormula}) and (\ref{GapFormula2})).  This is a crucial advantage in practical applications, since the short-range behavior of the pair correlation functions is more easily accessible in Quantum Monte Carlo simulation and less sensitive to finite size effects than the small-$q$ behavior of the structure factor, which appears in the Bijl-Feynman expression~(\ref{BFDispersion}).  

On the other hand, because of its ``collective" character the elastic approximation (equivalent to QLCA)  predicts a spectral gap at all densities, even in the limit of quantum degenerate plasma where no gap is expected as discussed above.

This problem is not too severe in the present study, because the electron-hole system is expected to spontaneously form Cooper pairs -- thus abandoning the normal degenerate plasma phase -- even at very high density.  The antisymmetric structure factor of the paired phase vanishes as $q^2$,  as one can easily verify from BCS theory  \cite{Littlewood1995,Littlewood1996} yielding 
\be
\label{SqBCS}
S(q)=-q^2\frac{1}{8\pi n}\int_0^{\infty} dk k\,  \frac{[\xi(k)\Delta'(k)-\xi'(k)\Delta(k)]^2}{[\Delta(k)^2+\xi(k)^2]^2},
\ee
and this gives, via Eq.~(\ref{BFDispersion}), a gapped dispersion $\omega_{BF}(0)>0$, which is compared with  $\omega_{QCM}(0)$ and $\omega_{\chi}(0)$ in  Fig.~\ref{fig:f7}.   The only issue here is that the elastic approximation overestimates the gap, possibly by a large factor, as discussed in the previous section. Otherwise, our results are qualitatively correct.

The prediction of a spectral gap is more problematic in systems with purely repulsive interactions -- for example,  electron-electron bilayers. In such systems no Cooper pairs are expected to form in the degenerate high-density limit, and therefore the system should remain gapless. This  contradicts the prediction of the QCM in the elastic approximation.   Nevertheless, it is quite possible that a gap, qualitatively similar to what QCM predicts, will appear  when the density is lowered below a certain critical value, at which the system enters a strongly correlated phase.  This scenario is strongly suggested by the classical simulations of Ref.~\onlinecite {kalman-ee} which show the existence of a gapped out-of-phase mode in the strongly correlated classical electron-electron system.  We may expect that the gap of the classical collective mode will manifest as a gap in the excitation spectrum of the corresponding quantum system.  If this expectation is correct, then a quantum phase transition must occur at some critical density, separating the gapless phase on the high-density side from the gapful phase on the low-density side. However, we find no evidence of such transition in the present theory.

One way to improve our treatment would be to modify the elastic approximation in such a way that it satisfies {\it both} the third-moment sum rule and the $S$-sum rule.   How to do this in a controlled manner remains an interesting question for further study.

 \section {Summary and Outlook}\label{sumup}
  
Let us recapitulate our main findings.

1. The elastic approximation of quantum continuum mechanics, applied to electron-hole bilayers, is equivalent to QLCA: this provides insight into the physical significance as well as the limits of validity of that approximation.\\

2. Like QLCA the theory predicts a gapped collective mode in the antisymmetric channel (i.e., when the carriers in the two layers oscillate with opposite phases). The frequency of this mode remains finite in the $q\to0$ limit and this finite value, denoted by $\omega(0)$, is ``the gap".\\    

3. The gap is found to exist at all densities, even though it becomes vanishingly small in the high-density limit as $h_{eh}(r)$ tends to $0$ (see Eq.~(\ref{GapFormula2})).  For the electron-hole bilayer this prediction makes perfect physical sense. Indeed, the  BCS mean field theory applied to this system predicts  the formation of Cooper pairs of electrons and holes with exponentially small binding energy at high density, evolving to exciton pairs at low density.  This is at variance with electron-electron bilayers, where the ordinary Fermi liquid phase is expected to be stable at high density.\\

4. A direct consequence of the gap, which has no equivalent in classical bilayers, is the incompressibility of the ground state, by which we mean the vanishing of the static density response function $\chi_-(q)$ as $q^2$ for $q\to0$.  The suppression of $\chi_-(q)$ at small $q$ is much stronger than the suppression expected from the RPA for a compressible electron gas ($\chi_{RPA}(q) \sim q$), and reflects the finite binding energy of electron-hole pairs.\\

5. The values of $\omega(0)$ calculated from the electron-hole pair distribution function of the Quantum Monte Carlo are in good agreement with those obtained from the pair distribution function of the BCS-like mean field theory.  They are also in good agreement with the values obtained by comparing the BCS density response function with the density response function obtained in QCM.  However, in this case, the good agreement breaks down in the limit of zero interlayer separation because QCM predicts a logarithmic divergence of $\omega(0)$ for $d\to 0$ which would make the density response function vanishes faster than $q^2$ for $q\to0$.\\

6. Our findings show that the QCM gap $\omega(0)$ calculated from a reliable set of pair distribution functions can be meaningfully related to the binding energy of bound states (excitons) for sufficiently large $d$ and $r_s$.\\

7. Incompressibility implies that the chemical potential of an electron has a discontinuous jump as a function of electron density $n_e$, when $n_e$ crosses the hole density $n_h$. The jump in chemical potential equals  the binding energy and can be measured experimentally by monitoring the electron and hole densities as functions of the voltages applied to the gates, as discussed in Ref.~\cite{Zeng2020}.\\

One drawback of our approach (QCM or QLCA) is that it predicts a gap in the out-of-phase mode at all densities {\it even in systems with only repulsive interactions} such as, for example, the {\it electron-electron} bilayer. This is conflict with the expectation that such systems should be  ordinary Fermi liquids at high density.  Physically, we do not expect the elastic approximation to work well at high density, because the excitation spectrum  (in the antisymmetric channel) becomes increasingly single-particle-like.  On the other hand, the low-density/strongly correlated limit should be described correctly if we assume that the electrons in the two layers freeze into two interlocking Wigner-crystal structures.  In the case of electron-hole bilayer this `correlated regime may set in even at high density, due to the occurrence of the Cooper instability. 
It has not been possible so far to formulate a  criterion to determine at what density, if any, the gap predicted by the present theory would cease to be reliable. We speculate that, with such criterion in hand, one could identify a quantum phase transition between a gapless phase at high density and a gapful one at low density.

 \section{Acknowledgments}
 GV acknowledges support for this project from the US Department of Energy (Office of Science) under grant No. DE-FG02-05ER46203.

\appendix 
\begin{subappendices}
\begin{widetext}
\section{Elastic approximation for a many-component system}\label{Appendix1}

In this section we generalize the elastic treatment  for a one-component system found in Gao~\cite{Gao10}  to a generic many-component system in dimension $D$. 
We start from the potential energy (see eq. (55) of \cite{Gao10}) duly generalized to a many-component system
 \be
W[{\bf u}]=\frac{1}{2}\sum_{\alpha,\beta}\int d{\bf r}\int d{\bf r}'\rho_{\alpha\beta}^{(2)}({\bf r},{\bf r}')\phi_{\alpha\beta}(|{\bf r}+{\bf u}_{\alpha}({\bf r})-{\bf r}'-{\bf u}_{\beta}({\bf r}'|),
 \ee
with $\phi{}_{\alpha\beta}(r)$ the interparticle potentials, $\rho_{\alpha\beta}^{(2)}({\bf r},{\bf r}')$
the two-body densities, and ${\bf u}_{\alpha}({\bf r})$ the displacement
field of the species $\alpha$. Below we shall denote with  ${u}_{\alpha;\mu}({\bf r})$  the cartesian component $\mu$ of  the $D-$dimensional vector  ${\bf u}_{\alpha}({\bf r})$. Following Gao we expand the potential
energy in powers of the displacement fields, to second order, obtaining
 \be
W_{2}[{\bf u}]=-\frac{1}{2}\sum_{\alpha,\beta;\mu,\nu}\frac{1}{2}\int d{\bf r}\int d{\bf r}'[K_{\alpha\beta}({\bf r},{\bf r}')]_{\mu\nu}  [u_{\alpha;\mu}({\bf r})-u_{\beta;\mu}({\bf r}')][u_{\alpha;\nu}({\bf r})-u_{\beta;\nu}({\bf r}')]\text{} \label{W2}
 \ee
where
 \be
[K_{\alpha\beta}  ({\bf r},{\bf r}')]_{\mu\nu}=\rho_{\alpha\beta}^{(2)}({\bf r},{\bf r}')\frac{\partial^{2}\phi_{\alpha\beta}(|{\bf r}-{\bf r}'|)}{\partial r_{\mu}\partial r'_{\nu}}.
 \ee
In an homogeneous  isotropic liquid
 
 \be
\rho_{\alpha\beta}^{(2)}({\bf r},{\bf r}')=\rho_{\alpha}\rho_{\beta}g_{\alpha\beta}(|{\bf r}-{\bf r}'|)=\rho_{\alpha\beta}^{(2)}({\bf |r}-{\bf r}'|)
 \ee
and therefore 
 \be
[K_{\alpha\beta}({\bf r}-{\bf r}')]_{\mu\nu}=\rho_{\alpha\beta}^{(2)}(|{\bf r}-{\bf r}'|)\frac{\partial^{2}\phi_{\alpha\beta}(|{\bf r}-{\bf r}'|)}{\partial r_{\mu}\partial r'_{\nu}}.
 \ee
We then use the definition of partial structure factors
 \be
S_{\alpha\beta}(q)=\delta_{\alpha\beta}+\sqrt{\rho_{\alpha}\rho_{\beta}}\int d{\bf r}[g_{\alpha\beta}(r)-1]\exp(i{\bf q}\cdot{\bf r}),
 \ee
or equivalently
 \be
S_{\alpha\beta}(q)=\delta_{\alpha\beta}+\sqrt{\rho_{\alpha}\rho_{\beta}}\int d{\bf r}g_{\alpha\beta}(r)\exp(i{\bf q}\cdot{\bf r})-(2\pi)^{D}\sqrt{\rho_{\alpha}\rho_{\beta}}\delta({\bf q}),
 \ee
to get the Fourier transform  of the two-body densities
 \be
\rho_{\alpha\beta}^{(2)}(q)=\sqrt{\rho_{\alpha}\rho_{\beta}}\left[S_{\alpha\beta}(q)-\delta_{\alpha\beta}+(2\pi)^{D}\sqrt{\rho_{\alpha}\rho_{\beta}}\delta({\bf q})\right].
 \ee
This yields, for  the Fourier transform of $[K_{\alpha\beta}(r)]_{\mu\nu}$,  
\ber
[K_{\alpha\beta}({\bf q})]_{\mu\nu}&=&\sqrt{\rho_{\alpha}\rho_{\beta}}\int\frac{d{\bf q}'}{(2\pi)^{D}}\left[S_{\alpha\beta}(|{\bf q}-{\bf q}'|)-\delta_{\alpha\beta}+(2\pi)^{D}\sqrt{\rho_{\alpha}\rho_{\beta}}\delta({\bf q}-{\bf q}')\right]q'_{\mu}q'_{\nu} \tilde{\phi}_{\alpha\beta}(q')\nonumber\\
&=&\sqrt{\rho_{\alpha}\rho_{\beta}}\int\frac{d{\bf q}'}{(2\pi)^{D}}\left[S_{\alpha\beta}(|{\bf q}-{\bf q}'|)-\delta_{\alpha\beta}\right]q'_{\mu}q'_{\nu} \tilde{\phi}_{\alpha\beta}(q')+\rho_{\alpha}\rho_{\beta}q_{\mu}q_{\nu} \tilde{\phi}_{\alpha\beta}(q)\label{K2}.
\eer
To write the equation of motions for the displacement fields we  need 
 \be
X_{\gamma;\lambda}(\mathbf{q})=\int d{\bf r}\frac{\partial W_{2}[{\bf u}]}{\partial u_{\gamma;\lambda}({\bf r})}\exp(i{\bf q}\cdot{\bf r}).
 \ee
From eq. (\ref{W2}) we get 
\be
\frac{\partial W_{2}[{\bf u}]}{\partial u_{\gamma;\lambda}({\bf r})}=-\sum_{\beta;\nu}\int d{\bf r}'K_{\gamma\beta;\lambda\nu}(|{\bf r}-{\bf r}'|)[u_{\gamma;\nu}({\bf r})-u_{\beta;\nu}({\bf r}')],
 \ee
which, combined with eq. (\ref{K2}), yields 
 \ber
X_{\gamma;\lambda}(\mathbf{q})&=&-\sum_{\beta;\nu}\left\{[K_{\gamma\beta}]_{\lambda\nu}({\bf q}=0)u_{\gamma;\nu}({\bf q})-[K_{\gamma\beta}]_{\lambda\nu}({\bf q})u_{\beta;\nu}({\bf q})\right\}\\
  & =&\sum_{\beta;\nu}\left[\left(-\sqrt{\rho_{\gamma}\rho_{\beta}}\int\frac{d{\bf q}'}{(2\pi)^{D}}\left[S_{\gamma\beta}(q')-\delta_{\gamma\beta}\right]q'_{\lambda}q'_{\nu} \tilde{\phi}_{\gamma\beta}(q')-\rho_{\gamma}\rho_{\beta}q_{\lambda}q_{\nu} \tilde{\phi}_{\gamma\beta}(q)|_{q=0}\right)u_{\gamma;\nu}({\bf q})\right.\nonumber \\
 && \left.\left(\sqrt{\rho_{\gamma}\rho_{\beta}}\int\frac{d{\bf q}'}{(2\pi)^{D}}\left[S_{\gamma\beta}(|{\bf q}-{\bf q}'|)-\delta_{\gamma\beta}\right]q'_{\lambda}q'_{\nu} \tilde{\phi}_{\gamma\beta}(q')+\rho_{\gamma}\rho_{\beta}q_{\lambda}q_{\nu} \tilde{\phi}_{\gamma\beta}(q)\right)u_{\beta;\nu}({\bf q)}\right].\label{eq:pot-0}
\eer

For the kinetic energy term using eqs. (27), (52),(53) of \cite{Gao10}
we obtain   for a homogeneus Fermion system  in dimension $D$
\begin{equation}
Y_{\gamma;\lambda}(\mathbf{q})=\frac{\partial T{}_{2}[{\bf u}]}{\partial\mathbf{u}_{\gamma}({\bf q})}=\frac{2}{D}\rho_{\gamma}t(\rho_{\gamma})\left[2{q_{\lambda}(\mathbf{q\cdot u}_{\gamma}(\mathbf{q}))+q^{2}\mathbf{u}_{\gamma}(\mathbf{q})}\right]+\frac{\rho_{\gamma}}{4m_{\gamma}}\hbar^{2}q^{2}{q_{\lambda}}(\mathbf{q\cdot}\mathbf{u_{\gamma}}(\mathbf{q}))\mathbf{},\label{eq:kin-1}
\end{equation}
with  $m_{\gamma}$  the  mass and $t(\rho_{\gamma})$ the interacting  kinetic energy per particle of the species $\gamma$. 
 The equation of motion for the species $\gamma$ thus becomes
 \be
m_{\gamma}\rho_{\gamma}\omega^{2}(\mathbf{q})\mathbf{u}_{\gamma}(\mathbf{q})=\mathbf{Y}_{\gamma}(\mathbf{q})+\mathbf{X}_{\gamma}(\mathbf{q}),
 \ee

with $\text{\ensuremath{\mathbf{Y}_{\gamma}}(\ensuremath{\mathbf{q}})}$
and $\mathbf{X}_{\gamma}(\mathbf{q})$  vectors  with cartesian components  $X_{\gamma;\lambda}(\mathbf{q})$ and $Y_{\gamma;\lambda}(\mathbf{q})$, i.e.,
\begin{align}
\mathbf{X}_{\gamma}(\mathbf{q}) & =\sum_{\beta}\left[\left(-\sqrt{\rho_{\gamma}\rho_{\beta}}\int\frac{d{\bf q}'}{(2\pi)^{D}}\left[S_{\gamma\beta}(q')-\delta_{\gamma\beta}\right]\mathbf{} \tilde{\phi}_{\gamma\beta}(q')\mathbf{q}'(\mathbf{q}\cdot'\mathbf{u}_{\gamma}({\bf q}))-\rho_{\gamma}\rho_{\beta} \tilde{\phi}_{\gamma\beta}(q_{0})\mathbf{q}_{0}(\mathbf{q}_{0}\cdot\mathbf{u}_{\gamma}({\bf q}))\left.\right|_{q_{0}=0}\right)\right.\nonumber \\
 & \left.+\left(\sqrt{\rho_{\gamma}\rho_{\beta}}\int\frac{d{\bf q}'}{(2\pi)^{D}}\left[S_{\gamma\beta}(|{\bf q}-{\bf q}'|)-\delta_{\gamma\beta}\right]\mathbf{ \tilde{\phi}_{\gamma\beta}(\mathnormal{q}')q'}(\mathbf{q}'\cdot\mathbf{u}_{\beta}({\bf q}))+\rho_{\gamma}\rho_{\beta} \tilde{\phi}_{\gamma\beta}(q)\mathbf{q}(\mathbf{q}\cdot\mathbf{u}_{\beta}({\bf q}))\right)\right],\label{eq:pot}
\end{align}
and 
\be
\mathbf{Y}_{\gamma}(\mathbf{q})=\frac{\partial T{}_{2}[{\bf u}]}{\partial\mathbf{u}_{\gamma}({\bf q})}=\frac{2}{D}\rho_{\gamma}t(\rho_{\gamma})\left[2{{\bf q}(\mathbf{q\cdot u}_{\gamma}(\mathbf{q}))+q^{2}\mathbf{u}_{\gamma}(\mathbf{q})}\right]+\frac{\rho_{\gamma}}{4m_{\gamma}}\hbar^{2}q^{2}{{\bf q}}(\mathbf{q\cdot}\mathbf{u_{\gamma}}(\mathbf{q}))\mathbf{}.
\ee

 \end{widetext}
\end{subappendices}

\bibliography{collective_bib}

\begin{thebibliography}{26}%
\makeatletter
\providecommand \@ifxundefined [1]{%
 \@ifx{#1\undefined}
}%
\providecommand \@ifnum [1]{%
 \ifnum #1\expandafter \@firstoftwo
 \else \expandafter \@secondoftwo
 \fi
}%
\providecommand \@ifx [1]{%
 \ifx #1\expandafter \@firstoftwo
 \else \expandafter \@secondoftwo
 \fi
}%
\providecommand \natexlab [1]{#1}%
\providecommand \enquote  [1]{``#1''}%
\providecommand \bibnamefont  [1]{#1}%
\providecommand \bibfnamefont [1]{#1}%
\providecommand \citenamefont [1]{#1}%
\providecommand \href@noop [0]{\@secondoftwo}%
\providecommand \href [0]{\begingroup \@sanitize@url \@href}%
\providecommand \@href[1]{\@@startlink{#1}\@@href}%
\providecommand \@@href[1]{\endgroup#1\@@endlink}%
\providecommand \@sanitize@url [0]{\catcode `\\12\catcode `\$12\catcode
  `\&12\catcode `\#12\catcode `\^12\catcode `\_12\catcode `\%12\relax}%
\providecommand \@@startlink[1]{}%
\providecommand \@@endlink[0]{}%
\providecommand \url  [0]{\begingroup\@sanitize@url \@url }%
\providecommand \@url [1]{\endgroup\@href {#1}{\urlprefix }}%
\providecommand \urlprefix  [0]{URL }%
\providecommand \Eprint [0]{\href }%
\providecommand \doibase [0]{http://dx.doi.org/}%
\providecommand \selectlanguage [0]{\@gobble}%
\providecommand \bibinfo  [0]{\@secondoftwo}%
\providecommand \bibfield  [0]{\@secondoftwo}%
\providecommand \translation [1]{[#1]}%
\providecommand \BibitemOpen [0]{}%
\providecommand \bibitemStop [0]{}%
\providecommand \bibitemNoStop [0]{.\EOS\space}%
\providecommand \EOS [0]{\spacefactor3000\relax}%
\providecommand \BibitemShut  [1]{\csname bibitem#1\endcsname}%
\let\auto@bib@innerbib\@empty
\bibitem [{\citenamefont {De~Palo}\ \emph {et~al.}(2002)\citenamefont
  {De~Palo}, \citenamefont {Rapisarda},\ and\ \citenamefont
  {Senatore}}]{depalo2002}%
  \BibitemOpen
  \bibfield  {author} {\bibinfo {author} {\bibfnamefont {S.}~\bibnamefont
  {De~Palo}}, \bibinfo {author} {\bibfnamefont {F.}~\bibnamefont {Rapisarda}},
  \ and\ \bibinfo {author} {\bibfnamefont {G.}~\bibnamefont {Senatore}},\
  }\href {\doibase 10.1103/PhysRevLett.88.206401} {\bibfield  {journal}
  {\bibinfo  {journal} {Phys. Rev. Lett.}\ }\textbf {\bibinfo {volume} {88}},\
  \bibinfo {pages} {206401} (\bibinfo {year} {2002})}\BibitemShut {NoStop}%
\bibitem [{\citenamefont {Senatore}\ and\ \citenamefont
  {Palo}(2003)}]{depalo2003}%
  \BibitemOpen
  \bibfield  {author} {\bibinfo {author} {\bibfnamefont {G.}~\bibnamefont
  {Senatore}}\ and\ \bibinfo {author} {\bibfnamefont {S.~D.}\ \bibnamefont
  {Palo}},\ }\href {\doibase 10.1002/ctpp.200310047} {\bibfield  {journal}
  {\bibinfo  {journal} {Contributions to Plasma Physics}\ }\textbf {\bibinfo
  {volume} {43}},\ \bibinfo {pages} {363} (\bibinfo {year} {2003})}\BibitemShut
  {NoStop}%
\bibitem [{\citenamefont {Shumway}\ and\ \citenamefont
  {Gilbert}(2012)}]{shumway2012}%
  \BibitemOpen
  \bibfield  {author} {\bibinfo {author} {\bibfnamefont {J.}~\bibnamefont
  {Shumway}}\ and\ \bibinfo {author} {\bibfnamefont {M.~J.}\ \bibnamefont
  {Gilbert}},\ }\href {\doibase 10.1103/PhysRevB.85.033103} {\bibfield
  {journal} {\bibinfo  {journal} {Phys. Rev. B}\ }\textbf {\bibinfo {volume}
  {85}},\ \bibinfo {pages} {033103} (\bibinfo {year} {2012})}\BibitemShut
  {NoStop}%
\bibitem [{\citenamefont {Maezono}\ \emph {et~al.}(2013)\citenamefont
  {Maezono}, \citenamefont {L\'opez~R\'{\i}os}, \citenamefont {Ogawa},\ and\
  \citenamefont {Needs}}]{maezono2013}%
  \BibitemOpen
  \bibfield  {author} {\bibinfo {author} {\bibfnamefont {R.}~\bibnamefont
  {Maezono}}, \bibinfo {author} {\bibfnamefont {P.}~\bibnamefont
  {L\'opez~R\'{\i}os}}, \bibinfo {author} {\bibfnamefont {T.}~\bibnamefont
  {Ogawa}}, \ and\ \bibinfo {author} {\bibfnamefont {R.~J.}\ \bibnamefont
  {Needs}},\ }\href {\doibase 10.1103/PhysRevLett.110.216407} {\bibfield
  {journal} {\bibinfo  {journal} {Phys. Rev. Lett.}\ }\textbf {\bibinfo
  {volume} {110}},\ \bibinfo {pages} {216407} (\bibinfo {year}
  {2013})}\BibitemShut {NoStop}%
\bibitem [{\citenamefont {Sharma}\ \emph {et~al.}(2016)\citenamefont {Sharma},
  \citenamefont {Saini},\ and\ \citenamefont {Bahuguna}}]{sharma2016}%
  \BibitemOpen
  \bibfield  {author} {\bibinfo {author} {\bibfnamefont {R.~O.}\ \bibnamefont
  {Sharma}}, \bibinfo {author} {\bibfnamefont {L.~K.}\ \bibnamefont {Saini}}, \
  and\ \bibinfo {author} {\bibfnamefont {B.~P.}\ \bibnamefont {Bahuguna}},\
  }\href {\doibase 10.1103/PhysRevB.94.205435} {\bibfield  {journal} {\bibinfo
  {journal} {Phys. Rev. B}\ }\textbf {\bibinfo {volume} {94}},\ \bibinfo
  {pages} {205435} (\bibinfo {year} {2016})}\BibitemShut {NoStop}%
\bibitem [{tra()}]{tramonto}%
  \BibitemOpen
  \href@noop {} {}\bibinfo {note} {F. Tramonto, Stefania De Palo, Saverio
  Moroni and Gaetano Senatore; (unpublished)}\BibitemShut {NoStop}%
\bibitem [{\citenamefont {Eisenstein}\ and\ \citenamefont
  {{MacDonald}}(2004)}]{Eisenstein2004}%
  \BibitemOpen
  \bibfield  {author} {\bibinfo {author} {\bibfnamefont {J.~P.}\ \bibnamefont
  {Eisenstein}}\ and\ \bibinfo {author} {\bibfnamefont {A.~H.}\ \bibnamefont
  {{MacDonald}}},\ }\href {\doibase 10.1038/nature03081} {\bibfield  {journal}
  {\bibinfo  {journal} {Nature}\ }\textbf {\bibinfo {volume} {432}},\ \bibinfo
  {pages} {691} (\bibinfo {year} {2004})}\BibitemShut {NoStop}%
\bibitem [{\citenamefont {Tutuc}\ \emph {et~al.}(2004)\citenamefont {Tutuc},
  \citenamefont {Shayegan},\ and\ \citenamefont {Huse}}]{Tutuc2004}%
  \BibitemOpen
  \bibfield  {author} {\bibinfo {author} {\bibfnamefont {E.}~\bibnamefont
  {Tutuc}}, \bibinfo {author} {\bibfnamefont {M.}~\bibnamefont {Shayegan}}, \
  and\ \bibinfo {author} {\bibfnamefont {D.~A.}\ \bibnamefont {Huse}},\ }\href
  {\doibase 10.1103/PhysRevLett.93.036802} {\bibfield  {journal} {\bibinfo
  {journal} {Phys. Rev. Lett.}\ }\textbf {\bibinfo {volume} {93}},\ \bibinfo
  {pages} {036802} (\bibinfo {year} {2004})}\BibitemShut {NoStop}%
\bibitem [{\citenamefont {Conti}\ \emph {et~al.}(2020)\citenamefont {Conti},
  \citenamefont {Van~der Donck}, \citenamefont {Perali}, \citenamefont
  {Peeters},\ and\ \citenamefont {Neilson}}]{Conti2020-I}%
  \BibitemOpen
  \bibfield  {author} {\bibinfo {author} {\bibfnamefont {S.}~\bibnamefont
  {Conti}}, \bibinfo {author} {\bibfnamefont {M.}~\bibnamefont {Van~der
  Donck}}, \bibinfo {author} {\bibfnamefont {A.}~\bibnamefont {Perali}},
  \bibinfo {author} {\bibfnamefont {F.~M.}\ \bibnamefont {Peeters}}, \ and\
  \bibinfo {author} {\bibfnamefont {D.}~\bibnamefont {Neilson}},\ }\href
  {\doibase 10.1103/PhysRevB.101.220504} {\bibfield  {journal} {\bibinfo
  {journal} {Phys. Rev. B}\ }\textbf {\bibinfo {volume} {101}},\ \bibinfo
  {pages} {220504} (\bibinfo {year} {2020})}\BibitemShut {NoStop}%
\bibitem [{\citenamefont {Van~der Donck}\ \emph {et~al.}(2020)\citenamefont
  {Van~der Donck}, \citenamefont {Conti}, \citenamefont {Perali}, \citenamefont
  {Hamilton}, \citenamefont {Partoens}, \citenamefont {Peeters},\ and\
  \citenamefont {Neilson}}]{Conti2020-II}%
  \BibitemOpen
  \bibfield  {author} {\bibinfo {author} {\bibfnamefont {M.}~\bibnamefont
  {Van~der Donck}}, \bibinfo {author} {\bibfnamefont {S.}~\bibnamefont
  {Conti}}, \bibinfo {author} {\bibfnamefont {A.}~\bibnamefont {Perali}},
  \bibinfo {author} {\bibfnamefont {A.~R.}\ \bibnamefont {Hamilton}}, \bibinfo
  {author} {\bibfnamefont {B.}~\bibnamefont {Partoens}}, \bibinfo {author}
  {\bibfnamefont {F.~M.}\ \bibnamefont {Peeters}}, \ and\ \bibinfo {author}
  {\bibfnamefont {D.}~\bibnamefont {Neilson}},\ }\href {\doibase
  10.1103/PhysRevB.102.060503} {\bibfield  {journal} {\bibinfo  {journal}
  {Phys. Rev. B}\ }\textbf {\bibinfo {volume} {102}},\ \bibinfo {pages}
  {060503} (\bibinfo {year} {2020})}\BibitemShut {NoStop}%
\bibitem [{\citenamefont {Wang}\ \emph {et~al.}(2019)\citenamefont {Wang},
  \citenamefont {Rhodes}, \citenamefont {Watanabe}, \citenamefont {Taniguchi},
  \citenamefont {Hone}, \citenamefont {Shan},\ and\ \citenamefont
  {Mak}}]{Mak2019}%
  \BibitemOpen
  \bibfield  {author} {\bibinfo {author} {\bibfnamefont {Z.}~\bibnamefont
  {Wang}}, \bibinfo {author} {\bibfnamefont {D.~A.}\ \bibnamefont {Rhodes}},
  \bibinfo {author} {\bibfnamefont {K.}~\bibnamefont {Watanabe}}, \bibinfo
  {author} {\bibfnamefont {T.}~\bibnamefont {Taniguchi}}, \bibinfo {author}
  {\bibfnamefont {J.~C.}\ \bibnamefont {Hone}}, \bibinfo {author}
  {\bibfnamefont {J.}~\bibnamefont {Shan}}, \ and\ \bibinfo {author}
  {\bibfnamefont {K.~F.}\ \bibnamefont {Mak}},\ }\href {\doibase
  10.1038/s41586-019-1591-7} {\bibfield  {journal} {\bibinfo  {journal}
  {Nature}\ }\textbf {\bibinfo {volume} {574}},\ \bibinfo {pages} {76}
  (\bibinfo {year} {2019})}\BibitemShut {NoStop}%
\bibitem [{\citenamefont {Vignale}\ and\ \citenamefont
  {MacDonald}(1996)}]{Vignale96}%
  \BibitemOpen
  \bibfield  {author} {\bibinfo {author} {\bibfnamefont {G.}~\bibnamefont
  {Vignale}}\ and\ \bibinfo {author} {\bibfnamefont {A.~H.}\ \bibnamefont
  {MacDonald}},\ }\href {\doibase 10.1103/PhysRevLett.76.2786} {\bibfield
  {journal} {\bibinfo  {journal} {Phys. Rev. Lett.}\ }\textbf {\bibinfo
  {volume} {76}},\ \bibinfo {pages} {2786} (\bibinfo {year}
  {1996})}\BibitemShut {NoStop}%
\bibitem [{\citenamefont {Santoro}\ and\ \citenamefont
  {Giuliani}(1988)}]{Santoro1988}%
  \BibitemOpen
  \bibfield  {author} {\bibinfo {author} {\bibfnamefont {G.~E.}\ \bibnamefont
  {Santoro}}\ and\ \bibinfo {author} {\bibfnamefont {G.~F.}\ \bibnamefont
  {Giuliani}},\ }\href {\doibase 10.1103/PhysRevB.37.937} {\bibfield  {journal}
  {\bibinfo  {journal} {Phys. Rev. B}\ }\textbf {\bibinfo {volume} {37}},\
  \bibinfo {pages} {937} (\bibinfo {year} {1988})}\BibitemShut {NoStop}%
\bibitem [{\citenamefont {Gao}\ \emph {et~al.}(2010)\citenamefont {Gao},
  \citenamefont {Tao}, \citenamefont {Vignale},\ and\ \citenamefont
  {Tokatly}}]{Gao10}%
  \BibitemOpen
  \bibfield  {author} {\bibinfo {author} {\bibfnamefont {X.}~\bibnamefont
  {Gao}}, \bibinfo {author} {\bibfnamefont {J.}~\bibnamefont {Tao}}, \bibinfo
  {author} {\bibfnamefont {G.}~\bibnamefont {Vignale}}, \ and\ \bibinfo
  {author} {\bibfnamefont {I.~V.}\ \bibnamefont {Tokatly}},\ }\href {\doibase
  10.1103/PhysRevB.81.195106} {\bibfield  {journal} {\bibinfo  {journal} {Phys.
  Rev. B}\ }\textbf {\bibinfo {volume} {81}},\ \bibinfo {pages} {195106}
  (\bibinfo {year} {2010})}\BibitemShut {NoStop}%
\bibitem [{\citenamefont {Girvin}\ \emph {et~al.}(1986)\citenamefont {Girvin},
  \citenamefont {MacDonald},\ and\ \citenamefont {Platzman}}]{Girvin1986}%
  \BibitemOpen
  \bibfield  {author} {\bibinfo {author} {\bibfnamefont {S.~M.}\ \bibnamefont
  {Girvin}}, \bibinfo {author} {\bibfnamefont {A.~H.}\ \bibnamefont
  {MacDonald}}, \ and\ \bibinfo {author} {\bibfnamefont {P.~M.}\ \bibnamefont
  {Platzman}},\ }\href {\doibase 10.1103/PhysRevB.33.2481} {\bibfield
  {journal} {\bibinfo  {journal} {Phys. Rev. B}\ }\textbf {\bibinfo {volume}
  {33}},\ \bibinfo {pages} {2481} (\bibinfo {year} {1986})}\BibitemShut
  {NoStop}%
\bibitem [{\citenamefont {Kalman}\ \emph {et~al.}(1999)\citenamefont {Kalman},
  \citenamefont {Valtchinov},\ and\ \citenamefont {Golden}}]{kalman-qlca}%
  \BibitemOpen
  \bibfield  {author} {\bibinfo {author} {\bibfnamefont {G.}~\bibnamefont
  {Kalman}}, \bibinfo {author} {\bibfnamefont {V.}~\bibnamefont {Valtchinov}},
  \ and\ \bibinfo {author} {\bibfnamefont {K.~I.}\ \bibnamefont {Golden}},\
  }\href {\doibase 10.1103/PhysRevLett.82.3124} {\bibfield  {journal} {\bibinfo
   {journal} {Phys. Rev. Lett.}\ }\textbf {\bibinfo {volume} {82}},\ \bibinfo
  {pages} {3124} (\bibinfo {year} {1999})}\BibitemShut {NoStop}%
\bibitem [{\citenamefont {Donk\'o}\ \emph {et~al.}(2003)\citenamefont
  {Donk\'o}, \citenamefont {Kalman}, \citenamefont {Hartmann}, \citenamefont
  {Golden},\ and\ \citenamefont {Kutasi}}]{kalman-ee}%
  \BibitemOpen
  \bibfield  {author} {\bibinfo {author} {\bibfnamefont {Z.}~\bibnamefont
  {Donk\'o}}, \bibinfo {author} {\bibfnamefont {G.~J.}\ \bibnamefont {Kalman}},
  \bibinfo {author} {\bibfnamefont {P.}~\bibnamefont {Hartmann}}, \bibinfo
  {author} {\bibfnamefont {K.~I.}\ \bibnamefont {Golden}}, \ and\ \bibinfo
  {author} {\bibfnamefont {K.}~\bibnamefont {Kutasi}},\ }\href {\doibase
  10.1103/PhysRevLett.90.226804} {\bibfield  {journal} {\bibinfo  {journal}
  {Phys. Rev. Lett.}\ }\textbf {\bibinfo {volume} {90}},\ \bibinfo {pages}
  {226804} (\bibinfo {year} {2003})}\BibitemShut {NoStop}%
\bibitem [{\citenamefont {Golden}\ \emph {et~al.}(2005)\citenamefont {Golden},
  \citenamefont {Mahassen}, \citenamefont {Kalman}, \citenamefont {Senatore},\
  and\ \citenamefont {Rapisarda}}]{gs-longitudinal}%
  \BibitemOpen
  \bibfield  {author} {\bibinfo {author} {\bibfnamefont {K.~I.}\ \bibnamefont
  {Golden}}, \bibinfo {author} {\bibfnamefont {H.}~\bibnamefont {Mahassen}},
  \bibinfo {author} {\bibfnamefont {G.~J.}\ \bibnamefont {Kalman}}, \bibinfo
  {author} {\bibfnamefont {G.}~\bibnamefont {Senatore}}, \ and\ \bibinfo
  {author} {\bibfnamefont {F.}~\bibnamefont {Rapisarda}},\ }\href {\doibase
  10.1103/PhysRevE.71.036401} {\bibfield  {journal} {\bibinfo  {journal} {Phys.
  Rev. E}\ }\textbf {\bibinfo {volume} {71}},\ \bibinfo {pages} {036401}
  (\bibinfo {year} {2005})}\BibitemShut {NoStop}%
\bibitem [{\citenamefont {Golden}\ \emph {et~al.}(2006)\citenamefont {Golden},
  \citenamefont {Mahassen}, \citenamefont {Senatore},\ and\ \citenamefont
  {Rapisarda}}]{gs-transverse}%
  \BibitemOpen
  \bibfield  {author} {\bibinfo {author} {\bibfnamefont {K.~I.}\ \bibnamefont
  {Golden}}, \bibinfo {author} {\bibfnamefont {H.}~\bibnamefont {Mahassen}},
  \bibinfo {author} {\bibfnamefont {G.}~\bibnamefont {Senatore}}, \ and\
  \bibinfo {author} {\bibfnamefont {F.}~\bibnamefont {Rapisarda}},\ }\href
  {\doibase 10.1103/PhysRevE.74.056405} {\bibfield  {journal} {\bibinfo
  {journal} {Phys. Rev. E}\ }\textbf {\bibinfo {volume} {74}},\ \bibinfo
  {pages} {056405} (\bibinfo {year} {2006})}\BibitemShut {NoStop}%
\bibitem [{\citenamefont {Giuliani}\ and\ \citenamefont {Vignale}(2005)}]{GV}%
  \BibitemOpen
  \bibfield  {author} {\bibinfo {author} {\bibfnamefont {G.}~\bibnamefont
  {Giuliani}}\ and\ \bibinfo {author} {\bibfnamefont {G.}~\bibnamefont
  {Vignale}},\ }\href {\doibase 10.1017/CBO9780511619915} {\emph {\bibinfo
  {title} {Quantum Theory of the Electron Liquid}}}\ (\bibinfo  {publisher}
  {Cambridge University Press},\ \bibinfo {year} {2005})\BibitemShut {NoStop}%
\bibitem [{\citenamefont {Golden}\ and\ \citenamefont
  {Kalman}(2003)}]{golden-2003}%
  \BibitemOpen
  \bibfield  {author} {\bibinfo {author} {\bibfnamefont {K.~I.}\ \bibnamefont
  {Golden}}\ and\ \bibinfo {author} {\bibfnamefont {G.~J.}\ \bibnamefont
  {Kalman}},\ }\href {\doibase 10.1088/0305-4470/36/22/306} {\bibfield
  {journal} {\bibinfo  {journal} {Journal of Physics A: Mathematical and
  General}\ }\textbf {\bibinfo {volume} {36}},\ \bibinfo {pages} {5865}
  (\bibinfo {year} {2003})}\BibitemShut {NoStop}%
\bibitem [{\citenamefont {Zeng}\ and\ \citenamefont
  {MacDonald}(2020)}]{Zeng2020}%
  \BibitemOpen
  \bibfield  {author} {\bibinfo {author} {\bibfnamefont {Y.}~\bibnamefont
  {Zeng}}\ and\ \bibinfo {author} {\bibfnamefont {A.~H.}\ \bibnamefont
  {MacDonald}},\ }\href {\doibase 10.1103/PhysRevB.102.085154} {\bibfield
  {journal} {\bibinfo  {journal} {Phys. Rev. B}\ }\textbf {\bibinfo {volume}
  {102}},\ \bibinfo {pages} {085154} (\bibinfo {year} {2020})}\BibitemShut
  {NoStop}%
\bibitem [{\citenamefont {Mahan}(1981)}]{Mahan81}%
  \BibitemOpen
  \bibfield  {author} {\bibinfo {author} {\bibfnamefont {G.~D.}\ \bibnamefont
  {Mahan}},\ }\href@noop {} {\emph {\bibinfo {title} {Many Particle Physics}}}\
  (\bibinfo  {publisher} {Plenum Press},\ \bibinfo {year} {1981})\BibitemShut
  {NoStop}%
\bibitem [{\citenamefont {Zhu}\ \emph {et~al.}(1995)\citenamefont {Zhu},
  \citenamefont {Littlewood}, \citenamefont {Hybertsen},\ and\ \citenamefont
  {Rice}}]{Littlewood1995}%
  \BibitemOpen
  \bibfield  {author} {\bibinfo {author} {\bibfnamefont {X.}~\bibnamefont
  {Zhu}}, \bibinfo {author} {\bibfnamefont {P.~B.}\ \bibnamefont {Littlewood}},
  \bibinfo {author} {\bibfnamefont {M.~S.}\ \bibnamefont {Hybertsen}}, \ and\
  \bibinfo {author} {\bibfnamefont {T.~M.}\ \bibnamefont {Rice}},\ }\href
  {\doibase 10.1103/PhysRevLett.74.1633} {\bibfield  {journal} {\bibinfo
  {journal} {Phys. Rev. Lett.}\ }\textbf {\bibinfo {volume} {74}},\ \bibinfo
  {pages} {1633} (\bibinfo {year} {1995})}\BibitemShut {NoStop}%
\bibitem [{\citenamefont {Littlewood}\ and\ \citenamefont
  {Zhu}(1996)}]{Littlewood1996}%
  \BibitemOpen
  \bibfield  {author} {\bibinfo {author} {\bibfnamefont {P.~B.}\ \bibnamefont
  {Littlewood}}\ and\ \bibinfo {author} {\bibfnamefont {X.}~\bibnamefont
  {Zhu}},\ }\href {\doibase 10.1088/0031-8949/1996/t68/008} {\bibfield
  {journal} {\bibinfo  {journal} {Physica Scripta}\ }\textbf {\bibinfo {volume}
  {T68}},\ \bibinfo {pages} {56} (\bibinfo {year} {1996})}\BibitemShut
  {NoStop}%
\bibitem [{not()}]{nota1}%
  \BibitemOpen
  \href@noop {} {}\bibinfo {note} {The third moment some rule for a system with
  two types of particles can be immediately obtained from eq. (3.206) for the
  spin dependent third moments of an electron gas \cite{GV}; one changes the
  coulomb interaction into pair interactions dependent on the spin indexes, to
  be then then interpreted as type indexes. Note that the third moment here is
  denoted with $M(q)q^2$, whereas in \cite{GV} is denoted with
  $M(q)$.}\BibitemShut {Stop}%
\end{thebibliography}%
\bibliographystyle{apsrev4-1}

\end{document}